\documentclass[12pt]{article}

\usepackage{preamble}
\usepackage{packages}
\usepackage{breqn}
\usepackage{enumerate}
\usepackage[font=small,labelfont=bf]{caption}
\usepackage[left=20mm,right=20mm,top=30mm,bottom=30mm]{geometry}
\usepackage{pgfplots}
\usetikzlibrary{calc}
\usepackage{extarrows}

\def\ASCommentColor{DarkRed}
\newcommand{\as}[1]{{\color{\ASCommentColor} \it NOTE: {#1} -- Adar ~~}}

\definecolor{darkgray}{rgb}{0.33, 0.33, 0.33}

\newcommand{\tr}{\text{tr}}
\normalsize
\title{\Large Transition of Large $R$-Charge Operators on a Conformal Manifold}
\author{Adar Sharon\footnote{adar.sharon@weizmann.ac.il} }
\author{Masataka Watanabe\footnote{masataka.watanabe@weizmann.ac.il}}
\affil{\small Department of Particle Physics and Astrophysics,\par Weizmann Institute of Science, Rehovot 7610001, Israel}
\normalsize
\date{}

\allowdisplaybreaks[4]

\begin{document}

  \maketitle
     \abstract{
     \normalsize
    We study the transition between phases at large $R$-charge on a conformal manifold. These phases are characterized by the behaviour of the lowest operator dimension $\Delta(Q_R)$ for fixed and large $R$-charge $Q_R$.
    We focus, as an example, on the $D=3$, $\mathcal{N}=2$ Wess-Zumino model with cubic superpotential $W=XYZ+\frac{\tau}{6}(X^3+Y^3+Z^3)$, and compute $\Delta(Q_R,\tau)$ using the $\epsilon$-expansion
    in three interesting limits. In two of these limits the (leading order) result turns out to be 
    \begin{equation*}
    \Delta(Q_R,\tau)=
        \begin{cases}
         \left(\text{BPS bound}\right)\left[1+O(\epsilon \abs{\tau}^2Q_R)\right], & Q_R\ll \left\{ \frac{1}{\epsilon},\, \frac{1}{\epsilon\abs{\tau}^2}\right\}\\
         \frac{9}{8}\left(\frac{\epsilon\abs{\tau}^2}{2+\abs{\tau}^2}\right)^{\frac{1}{D-1}}Q_R^{\frac{D}{D-1}}
         \left[1+O\left(\left(\epsilon \abs{\tau}^2Q_R\right)^{-\frac{2}{D-1}}\right)\right],
         & Q_R\gg \left\{ \frac{1}{\epsilon},\, \frac{1}{\epsilon\abs{\tau}^2}\right\}
        \end{cases}
    \end{equation*}
    which leads us to the double-scaling parameter, $\epsilon \abs{\tau}^2Q_R$, which interpolates between the ``near-BPS phase'' ($\Delta(Q)\sim Q$) and the ``superfluid phase'' ($\Delta(Q)\sim Q^{D/(D-1)}$) at large $R$-charge.
    This smooth transition, happening near $\tau=0$, is a large-$R$-charge manifestation of the existence of a moduli space and an infinite chiral ring at $\tau=0$.
    We also argue that this behavior can be extended to three dimensions with minimal modifications, 
    and so we conclude that $\Delta(Q_R,\tau)$ experiences a smooth transition around $Q_R\sim 1/\abs{\tau}^2$.
    Additionally, we find a first-order phase transition for $\Delta(Q_R,\tau)$ as a function of $\tau$, as a consequence of the duality of the model.
    We also comment on the applicability of our result down to small $R$-charge.
      }
      
      \newpage
\normalsize
\setcounter{tocdepth}{2}
\setcounter{secnumdepth}{4}
\tableofcontents
\hypersetup{linkcolor=PaleGreen4}
\newpage
\normalsize

\section{Introduction}
\label{sec:intro}

The large global charge sector of generic conformal field theories is known to simplify \cite{Hellerman:2015nra,Alvarez-Gaume:2016vff,Monin:2016jmo}.
Even when the theory lacks any weak-coupling parameters (or even a Lagrangian description), one can compute dimensions and OPE coefficients to any given order in the inverse charge expansion for operators with a large global charge.

The reason why the large charge sector becomes simple can be understood from the effective theory at large chemical potential for the charge density $\rho$.
First, quantum fluctuations of the theory become negligible.
This is because adding such a chemical potential leads to a natural separation of scales, \it i.e., \rm the ratio between the UV and the IR scales are given by
$\Lambda_{\rm IR}/\Lambda_{\rm UV}\propto Q^{-\alpha}$ for some $\alpha>0$, where $Q$ denotes the global charge which we take large.
Furthermore, since the system is conformally invariant in the IR fixed point, the classical Weyl invariance (because we can ignore quantum corrections at large charge) constrains the form of the effective field theory (EFT) quite strongly. 
In many interesting examples, there are only a finite number of operators which can appear at any given order in the $1/Q$ expansion, which makes the problem tractable \cite{Hellerman:2015nra,Monin:2016jmo,Hellerman:2017veg,Hellerman:2017sur,Favrod:2018xov,Kravec:2018qnu,Hellerman:2018xpi,Hellerman:2020sqj}.

Quantum field theories exhibit several interesting phases at large global charge.
Even if we limit ourselves to the sector where only one Cartan of the global symmetry is large,\footnote{See \cite{Alvarez-Gaume:2016vff,Hellerman:2017efx,Hellerman:2018sjf,Loukas:2017hic} for cases where more than one Cartan are excited.} there have already appeared several different behaviours at large charge in the literature.\footnote{
This is to be contrasted with the large spin sector of theories in more than two dimensions, which always becomes generalised free \cite{1212.4103,1212.3616,1611.01500,1612.00696}, studied in the context of the analytic bootstrap.
An interesting direction of treating large charge and large spin simultaneously has also been pursued  \cite{Kravec:2019djc,Cuomo:2017vzg,Cuomo:2019ejv}.
See also \cite{Jafferis:2017zna} for an attempt to study the large charge sector using analytic bootstrap.
}
The most prominent examples are (we stress that these are certainly not all the possibilities):
\begin{enumerate}[(1)]
    \item {\bf BPS phase/Free phase}: The effective theory at leading order is just free, with the lowest operator dimension going as $\Delta(Q)\propto Q$. Examples include the trivial case of the theory of free complex boson, and supersymmetric theories with moduli spaces of vacua (see for example \cite{Hellerman:2017veg,Hellerman:2017sur,Hellerman:2018xpi,Hellerman:2020sqj}).\footnote{Strictly speaking, these two are in different universality classes, because in the SUSY case, fermions also become gapless. We neglect this difference hereafter, as it does not affect the scaling exponent of the lowest operator at large global charge.}
    \item {\bf Superfluid phase/Semi-classical phase}: The underlying theory is typically strongly coupled, and the lowest operator dimension goes as $\Delta(Q)\propto Q^{\frac{D}{D-1}}$.\footnote{Although this might seem to be in contradiction with the presence of SUSY, as the SUSY is spontaneously broken by the state we want to look at, this is not a problem. We will elaborate on this later on.} This includes the $O(2)$ Wilson-Fisher fixed-point in $D=3$ and the $D=3$, $\mathcal{N}=2$ supersymmetric Ising model \cite{Hellerman:2015nra}. We call this phase ``semi-classical'', because it is a consequence of the (almost) classical Weyl invariance of the EFT at large charge.
    \item {\bf Fermi sea}: The theory forms a Fermi sea at large charge. One example is the theory of free fermions, where the lowest operator dimension is $\Delta(Q)\propto Q^{\frac{d}{d-1}}$.
    Note that although the scaling here is the same as that of the superfluid phase, we have different gapless modes on top.
\end{enumerate}
In this paper we will mostly consider the BPS and the superfluid phases. We will only comment on possible directions involving the Fermi sea phase towards the end.

Since we will use the state-operator correspondence to compute the operator dimension from the EFT at large charge, we put the theory on the unit sphere, a finite volume spatial slice.
This means there are usually no real phase transitions between these phases, and we might be able to interpolate between some of them.\footnote{This is not completely true when we take $Q$ to be strictly infinite, but should be true at large but finite $Q$.} The Wilson-Fisher fixed-point in $D=4-\epsilon$, for example, has a weak coupling limit $\epsilon\to 0$ which interpolates phases (1) and (2) \cite{Arias-Tamargo:2019xld,Badel:2019oxl,Watanabe:2019pdh,Arias-Tamargo:2019kfr,Badel:2019khk,Antipin:2020abu,Antipin:2020rdw}.
More precisely, as was pointed out in those papers, the behaviour that interpolates them is controlled by a double-scaling parameter $\epsilon Q$, where large $\epsilon Q$ leads to phase (2), while small $\epsilon Q$ leading to phase (1).\footnote{
This might be a hint of why the large charge expansion numerically works even for $Q=O(1)$ \cite{Banerjee:2017fcx,Banerjee:2019jpw,delaFuente:2018qwv,Karthik:2018rcg}.
} Note that the infinite volume limit $Q\to \infty$ with $\epsilon$ fixed recovers the phase transition at $\epsilon=0$ from phase (2) to (1).

In this paper, we consider a similar transition for $\mathcal{N}=2$ supersymmetric (SUSY) theories at large $R$-charge.
Although we have supersymmetry, the operators we will be studying generically lie beyond the regime of SUSY protected operators in which they have been mostly studied.\footnote{
In fact, the double-scaling limit in the context of the large $R$-charge was first found for the $\mathcal{N}=2$ SUSY gauge theories, where the double-scaling parameter $g_{\rm YM}Q^2$ was introduced as a simplifying limit for computing correlation functions of BPS operators \cite{Bourget:2018obm,Beccaria:2018xxl,Beccaria:2018owt,Hellerman:2018xpi,Grassi:2019txd,Beccaria:2020azj}.
}
We will be looking at a theory with a conformal manifold on which the chiral ring truncates except at some special points.\footnote{This is a generic phenomenon for the types of Wess-Zumino theories we will consider \cite{Komargodski:2020ved}.}
Specifically, we will look at the $D=3$, $\mathcal{N}=2$ SUSY Wess-Zumino model with three chiral superfields $X,Y,Z$ and a cubic superpotential \begin{equation}
    W(X,Y,Z)=g\left(XYZ+\frac{\tau}{6} (X^3+Y^3+Z^3)\right)\;.
\end{equation}
This model (which incidentally is mirror dual to $\mathcal{N}=2$ SQED with one flavour deformed by some monopole superpotential, so that the analysis in this paper is applicable there too \cite{Aharony:1997bx,Intriligator:1996ex,deBoer:1997kr,Benini:2017dud}) is known to possess a conformal manifold parametrised by $\tau$, and is strongly-coupled on the entire conformal manifold \cite{Leigh:1995ep,Strassler:1998iz,Green:2010da,Kol:2002zt,Kol:2010ub}. 
The existence of the conformal manifold can be analytically continued to $4-\epsilon$ dimensions too \cite{Baggio:2017mas,Ferreira:1996ug,Ferreira:1996az}.
While the theory is strongly-coupled throughout the conformal manifold in three dimensions,
for small $\epsilon$ the coupling is $\abs{g}^2\propto {\epsilon}$ and thus the theory is weakly-coupled in the $\epsilon$-expansion.

On this conformal manifold, the chiral ring truncates except at $\tau=0$.\footnote{There are additional special values of $\tau$ which are equivalent to $\tau=0$ due to a duality, as will be explained soon.}
In terms of the EFT at large $R$-charge, when the chiral ring does not truncate, there is a moduli space of vacua and hence the leading order EFT becomes free (BPS phase). This means that the lowest operator dimension at large $R$-charge saturates the BPS bound, as expected \cite{Hellerman:2017veg}.
On the other hand, when the chiral ring truncates the leading term of the EFT becomes semi-classical (superfluid phase), and the lowest operator dimension at large $R$-charge $Q_R$ scales as $Q_R^{{D}/({D-1})}$, which is parametrically beyond the regime of SUSY protected dimensions.\footnote{For more information, see \cite{Hellerman:2015nra,Hellerman:2017veg}. In the former the case where $\tau=0$ was studied, while in the latter the $W=\Phi^3$ model was studied. Three copies of the latter corresponds to the $\tau=\infty$ point of our model.} 
Put another way, the fixed-charge state spontaneously breaks SUSY completely when $\tau\neq 0$, so that the lowest operator at fixed and large $R$-charge does not, or should not, saturate the BPS bound. On the other hand, at $\tau=0$ the fixed-charge state does not break (some parts of) the SUSY algebra, and hence the lowest operator dimension should saturate the BPS bound.
As there are two distinct spontaneous symmetry breaking patterns and hence two different phases at large $R$-charge,
we can see that there must be a transition between the free phase and the superfluid phase as we vary $\tau$.
In the limit where we take the $R$-charge to infinity, this transition is a phase transition, which happens exactly at $\tau=0$.

For large but finite $R$-charge, this transition from phase (1) and (2) should happen near $\tau=0$, controlled by some double-scaling parameter as expected from the $O(2)$ Wilson-Fisher fixed point case.
One major difference in our case compared to the $O(2)$ Wilson-Fisher case is that the two phases are interpolated by full-fledged, unitary theories for any value of $\tau$ in three dimensions.\footnote{Note that this also happens for the $O(N)$ Wilson-Fisher theory at large-$N$ \cite{Alvarez-Gaume:2019biu}. Compare this with the large-$N$ limit of the $\mathbb{C}P^{N-1}$ model, where this does not happen \cite{Dyer:2015zha,delaFuente:2018qwv}.}
Nevertheless in order to find the double-scaling parameter, we resort to the $\epsilon$-expansion and compute the lowest operator dimension at large $R$-charge -- otherwise there are no weak-coupling parameters.\footnote{
We might as well analyse the theory using the EFT at large $R$-charge, without using the $\epsilon$-expansion.
However, as the EFT typically can only describe the regime in which $Q$ is the largest parameter in the system, this method might not be suited for the analysis of the transition, and thus we don't pursue it.
See \cite{Watanabe:2019pdh} for an attempt to avoid this situation.
}
We then partially resum the $\epsilon$-expansion using the input from the EFT, to extrapolate to $\epsilon=1$ and draw conclusions about what happens in three dimensions.

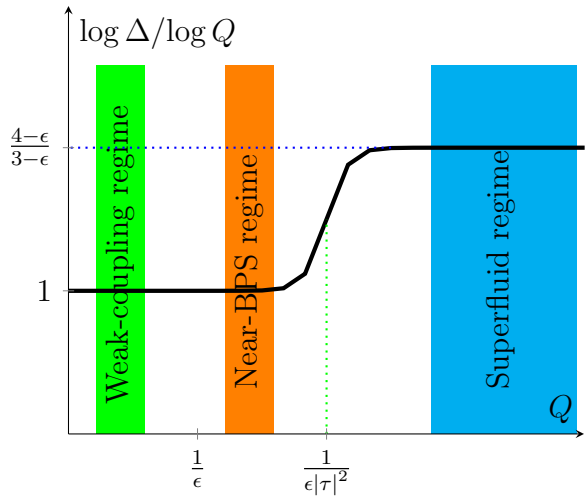
\begin{figure}[]
\begin{center}
    \begin{tikzpicture}
\begin{axis}[
    xmin=0, xmax=7.99,
    ymin=0, ymax=2.99,
    xtick distance=8,
    ytick distance=8,
    axis lines=center,
    axis on top=true,
    extra y ticks={1,2},
    extra y tick labels={$1$,$\frac{4-\epsilon}{3-\epsilon}$},
    extra x ticks={2,4},
    extra x tick labels={$\frac{1}{\epsilon}$,$\frac{1}{\epsilon\abs{\tau}^2}$},
    domain=0:8,
    ylabel=$\log\Delta/\log Q$,
    xlabel=$Q$,
    ]
    \node [fill=green,rotate=90] at (axis cs:0.8,1.2) {\hspace{5mm} Weak-coupling regime \hspace*{3mm}};
    \node [fill=orange,rotate=90] at (axis cs:2.8,1.13) {\hspace{14mm} \;Near-BPS regime \hspace*{15mm}};
    \node [fill=cyan,rotate=90] at (axis cs:6,1.2) {\hspace{7mm} \phantom{Superfluid regime} \hspace*{9mm}};
    \node [fill=cyan,rotate=90] at (axis cs:6.5,1.2) {\hspace{7mm} \phantom{Superfluid regime} \hspace*{9mm}};
    \node [fill=cyan,rotate=90] at (axis cs:7,1.2) {\hspace{7mm} \phantom{Superfluid regime} \hspace*{9mm}};
    \node [fill=cyan,rotate=90] at (axis cs:7.5,1.2) {\hspace{7mm} \phantom{Superfluid regime} \hspace*{9mm}};
    \node [fill=cyan,rotate=90] at (axis cs:6.7,1.2) {\hspace{7mm} {Superfluid regime} \hspace*{9mm}};
    \node [fill=white] at (axis cs:4,2.75) {\hspace{7mm} \phantom{\Huge aaaaaaaaaaaaaaaaaaaaaaaaaaaaaaaaaaaaaaaaaaaaaaaaaaaaaaaaaaaaaaaaaaaaaaaaaaaaaaaaaaaaaaaaaaaaaaaaaaaaaaaaaaaaaaaaaaaaaaaaaaaaaaaaaaaaaaaaaaaaa} \hspace*{9mm}};
    \addplot [mark=none,draw=black,ultra thick] {1.5+tanh((\x-4)*3)/2};
    \draw [blue, dotted, thick] (axis cs:5,2)-- (axis cs:0,2);
    \draw [green, dotted, thick] (axis cs:4,0)-- (axis cs:4,1.5);
\end{axis}
\end{tikzpicture}
    \caption{A schematic plot of the behaviour of the leading power of $Q$ in $\Delta(Q)$ at large $Q$, \it i.e., \rm $\log\Delta/\log Q$.
    The three interesting regions (superfluid, BPS, weak-coupling) correspond to three different regions in the $Q$ axis separated by the ticks.
    The smooth transition of the exponent happens around $Q=\frac{1}{\epsilon\abs{\tau}^2}$.}
    \label{fig:aa}
    \end{center}
\end{figure}

Several new features appear for the cubic model we are going to study, compared with the $O(2)$ Wilson-Fisher theory in $4-\epsilon$ dimensions.
First, as we already mentioned, one can observe a transition using only full-fledged unitary theories in $D=3$, without working in non-integer dimensions.
Second, as we have two parameters we can control and which we can take to be small, there are more than 
one candidates for the double-scaling parameter.
Because of this, we have computed the lowest operator dimension in three regions of interest:
\begin{itemize}
\item {\bf Superfluid Regime}: $1\ll\left\{ 1/\epsilon,\, 1/\left(\epsilon\abs{\tau}^2\right)\right\}\ll Q_R$ 
\item {\bf Near-BPS Regime}: $1\ll 1/\epsilon \ll Q_R\ll 1/\left(\epsilon\abs{\tau}^2\right)$
\item {\bf Weak-coupling Regime}: $1\ll Q_R\ll \left\{ 1/\epsilon,\, 1/\left(\epsilon\abs{\tau}^2\right)\right\}$
\end{itemize}
Note that the weak-coupling regime disappears when we extrapolate $\epsilon$ to $1$, and the superfluid and near-BPS regimes will be the transition of interest in three dimensions.\footnote{
In other words, the last regime cannot be consistently extrapolated to the strongly-coupled regime in three dimensions, whereas the first two can be extrapolated to three dimensions, which we will describe in the discussion section.
}
Also note that the near-BPS regime only exists when $\abs{\tau}^2\ll 1$.
We will show that as the name suggests, the theory is in the superfluid phase in the first regime, whereas in the second and the third regimes they are in the BPS/free phase --
The difference between the last two regimes is at subleading orders in the $\abs{\tau}^2 Q_R\ll 1$ expansion, so that we find the double-scaling parameter to be $\epsilon\abs{\tau}^2Q_R$.
We show the schematic picture of the double-scaling behaviour in Fig. \ref{fig:aa}.

Another important aspect of this theory, which was not present in the $O(2)$ Wilson-Fisher theory, is the presence of a duality \cite{Lerche:1989cs,Baggio:2017mas}.
This duality is a straightforward one, in that it is just UV duality, relating different theories \it via \rm a combination of the redefinition of the fields $X,Y,Z$ and a transformation of $\tau$.
Nevertheless this duality teaches us many things about the operator spectrum of the theory.
For instance, one can see that the form of the lowest operator dimension undergoes a first-order phase transition as we vary $\tau$.
This type of phenomenon, although a trivial consequence of this UV duality, has never been observed in other examples of double-scaling behaviour of theories at large charge and weak coupling --
For the $O(2)$ Wilson-Fisher theory in $4-\epsilon$ dimensions, the form of the lowest operator is $\phi^n$ irrespective of the double-scaling parameter $\epsilon n$, while for our theory, as we vary $\abs{\tau}^2 Q_R$, the form of the lowest operator changes suddenly as we move out of the fundamental domain of the conformal manifold determined by the duality group.
We will also discuss the specific form of the lowest-dimension operator at large charge later on, using the duality action on the conformal manifold.

The rest of this paper is organized as follows. In Section \ref{sec:D=3_cubic_superpotential} we review the $D=3$, $\mathcal{N}=2$ Wess-Zumino theory, along with its conformal manifold, its global symmetries and the duality. In Section \ref{sec:saddle_point} we set the groundwork for the large charge expansion, including finding the saddle point and discussing the role of the fermions. 
Next in Section \ref{sec:dimension} we compute the dimension of the lowest operators at large $R$-charge in the $\epsilon$-expansion, up to one-loop.
We resum the $\epsilon$-expansion using the input from the EFT in Section \ref{sec:analysis} to extrapolate the result to three dimensions. 
In Section \ref{sec:validity}, we discuss the range of validity of our computations and discuss the form of the lowest dimension operator at large $R$-charge. Specifically, we will argue that the extrapolation of our result down to small $R$-charge requires careful consideration of non-perturbative corrections in the inverse $R$-charge expansion.
Finally we summarize our results in Section \ref{sec:conclusion}. Some calculations required for the 1-loop contribution to the lowest operator dimension appear in Appendix \ref{sec:excitation}.

\section{The cubic superpotential model in $D=3$}\label{sec:D=3_cubic_superpotential}

\subsection{The model and its global symmetry}

The model we consider in this paper is the $D=3$ $\mathcal{N}=2$ supersymmetric Wess-Zumino (WZ) model on $S^2\times\mathbb{R}$ with the following K\"ahler potential $K$ and superpotential  $W$,
\begin{eqnarray}
K(X,Y,Z)&=&\abs{X}^2+\abs{Y}^2+\abs{Z}^2\nonumber\;,\\
W(X,Y,Z)&=&g\left(XYZ+\frac{\tau}{6} (X^3+Y^3+Z^3)\right)\;.\label{eq:theory}
\end{eqnarray}
Here $X$, $Y$, $Z$ are chiral superfields and $g$ and $\tau$ are the (generically complex) coupling constants.
Note that the phase of $g$ does not change physics at all, as the component Lagrangian only depends on $\abs{g}^2$.
We also set the radius of the sphere to be one, since we are only interested in the operator dimensions of the theory.
This theory is known to flow to an IR fixed-point for any value of $\tau\in\mathbb{C}$, and so it has a complex one-dimensional conformal manifold parametrised by $\tau$ \cite{Baggio:2017mas,Strassler:1998iz}.\footnote{One can also argue for the existence of this conformal manifold using the methods described in \cite{Green:2010da} (see also \cite{Kol:2002zt,Kol:2010ub}), or alternatively using the method in \cite{Sharon:2020doo}.} On a generic point of the conformal manifold, the global symmetry of this theory consists of a $U(1)_R$ symmetry as well as a finite group $G$ (which is a subgroup of $U(3)$ rotating $X$, $Y$ and $Z$) generated by \cite{Baggio:2017mas}
\begin{eqnarray}
\begin{pmatrix}
X \\
Y \\
Z 
\end{pmatrix}&\mapsto&
\begin{pmatrix}
1 & 0 &0 \\
0 & 0 & 1 \\
0 & 1 & 0 
\end{pmatrix}
\begin{pmatrix}
X \\
Y \\
Z 
\end{pmatrix}\\
\begin{pmatrix}
X \\
Y \\
Z 
\end{pmatrix}&\mapsto&
\begin{pmatrix}
0 & 1 & 0 \\
1 & 0 & 0 \\
0 & 0 & 1 
\end{pmatrix}
\begin{pmatrix}
X \\
Y \\
Z 
\end{pmatrix}\\
\begin{pmatrix}
X \\
Y \\
Z 
\end{pmatrix}&\mapsto&
\begin{pmatrix}
1 & 0 & 0 \\
0 & e^{i2\pi/3} & 0 \\
0 & 0 & e^{i4\pi/3} 
\end{pmatrix}
\begin{pmatrix}
X \\
Y \\
Z 
\end{pmatrix}
\end{eqnarray}
At the infrared fixed point, the chiral operators $X$, $Y$, and $Z$ all have $R$-charge $2/3$.
Because these operators saturate the BPS bound, their dimensions are $\frac{D-1}{2}$ times the $R$-charge, \it i.e. \rm $\frac{D-1}{3}$.
For later convenience we also define the $Q$-charge, which is related to the $R$-charge, $Q_R$, \it via \rm
\begin{equation}
Q=\frac{3}{2}Q_R,
\end{equation}
under which fundamental fields $X$, $Y$, and $Z$ all have charge $Q=1$.

\subsection{The duality}
\label{qq}

The theory with superpotential \eqref{eq:theory} enjoys a discrete duality group $G_d\subset U(3)$ which acts on the fields $X$, $Y$, and $Z$ \cite{Baggio:2017mas}. 
This duality can be found by noticing that the K\"ahler potential is invariant under $U(3)$, while the superpotential can be made invariant under a subgroup $G_d$ of $U(3)$ if we also assign a transformation to the exactly marginal coupling $\tau$. 
This means that different values of $\tau$ are related by a redefinition of $X$, $Y$, and $Z$, so that the theory should be the same.

More concretely, a generic element of $U(3)$ will transform the superpotential to the generic form $W=h_{ijk}X^i X^j X^k$, where $X=X_1$, $Y=X_2$ and $Z=X_3$. However, the elements of the duality group $G_d$ will preserve its form \eqref{eq:theory}, with a certain transformation rule for $\tau$ as well as ${g}$.
One can understand this as nothing but a symmetry of the UV action with $\tau$ being a background field.
$G_d$ can be described using three generators (see Table \ref{tab:duality-generator}). 
Incidentally, upon close inspection of this duality group one finds it is isomorphic to the symmetric group $S_4$.
\begin{table}[htbp]
\begin{center}
  \begin{tabular}{|c||c|c|} \hline
      Reparametrization & $\tau$ & $g$  \\ \hline\hline
    $\begin{pmatrix}
X \\
Y \\
Z 
\end{pmatrix}\mapsto
\begin{pmatrix}
\bar{X} \\
\bar{Y} \\
\bar{Z} 
\end{pmatrix}$ & $\tau\mapsto \bar\tau$ & $g\mapsto \bar g$ \\ \hline   
    $\begin{pmatrix}
X \\
Y \\
Z 
\end{pmatrix}\mapsto
\begin{pmatrix}
\omega^2{X} \\
{Y} \\
{Z} 
\end{pmatrix}$ & $\tau\mapsto \omega\tau$ & $g\mapsto  \omega^2g$  \\ \hline
    $\begin{pmatrix}
X \\
Y \\
Z 
\end{pmatrix}\mapsto\displaystyle\frac{1}{\sqrt{3}}
\begin{pmatrix}
1 & 1 & \omega^2 \\
1 & \omega & \omega \\
\omega & 1 & \omega 
\end{pmatrix}
\begin{pmatrix}
X \\
Y \\
Z 
\end{pmatrix}$ & $\displaystyle\tau\mapsto \frac{\tau+2 \omega}{\omega^2\tau -1}$ & $\displaystyle g\mapsto \frac{g(\omega\tau-1)}{\sqrt{3}}$  \\ \hline
  \end{tabular}
  \caption{Generators of the duality group $G_d$. Here $\omega=e^{2\pi i/3}$ is a cubic root of unity. There is a slight mismatch between this table and the one given in \cite{Baggio:2017mas}; we believe the results here are the correct ones. The difference does not affect any computation in either paper.}
  \label{tab:duality-generator}
\end{center}
\end{table}

\begin{figure}[htbp]
    \center{\includegraphics[width=0.5\textwidth]
    {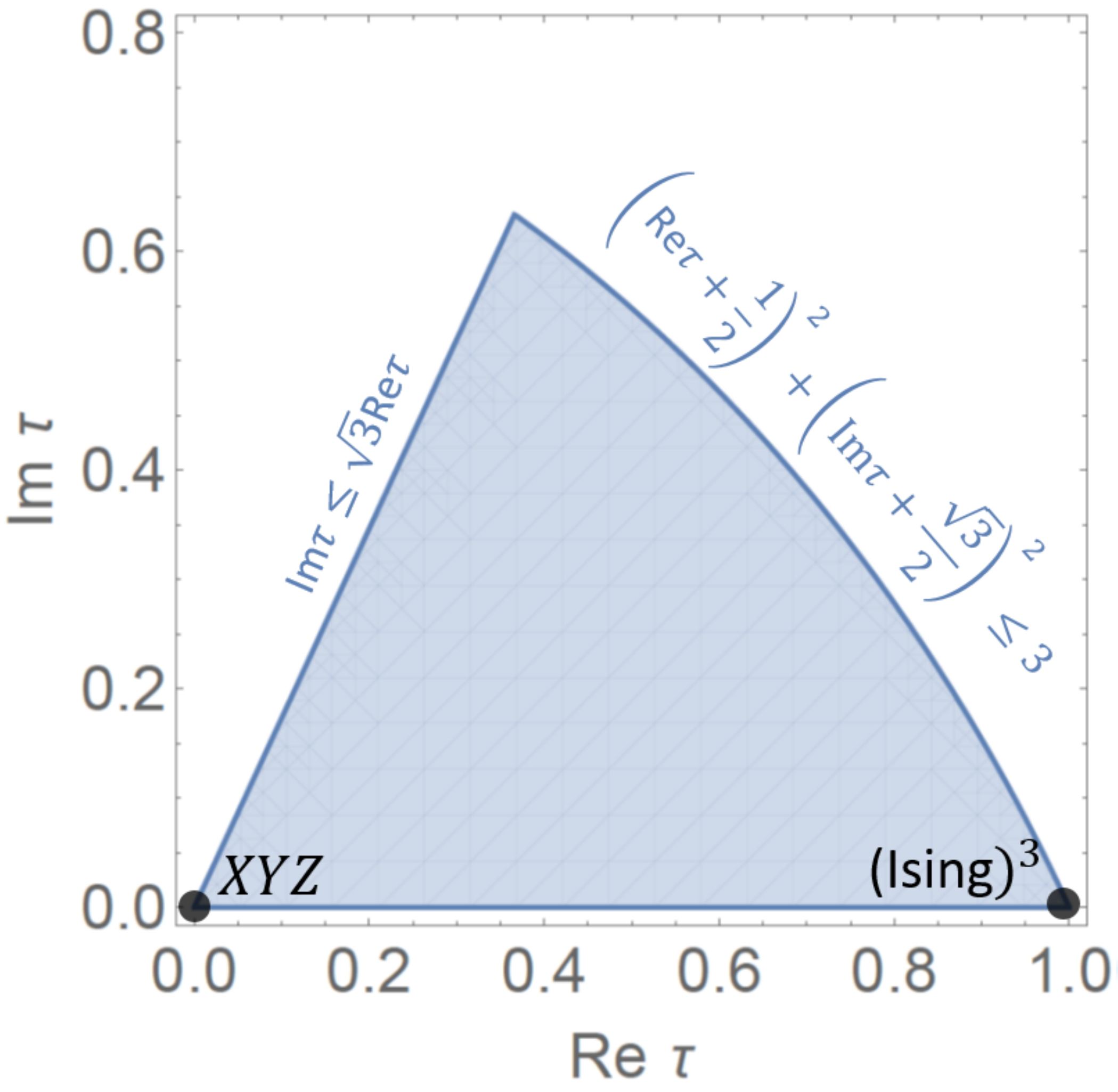}}
    \caption{The fundamental domain of the duality group $G_d$. On the boundary of the fundamental domain some of the duality action turns into the global symmetry. Two of the three corners of the region, marked in dots, correspond to previously well-studied theories, namely the $W=XYZ$ model and (three copies of) the $\mathcal{N}=2$ SUSY Ising model.}
    \label{fig:FundamentalDomain}
\end{figure}

One can now define the ``fundamental domain'' $\mathcal{M}\equiv\mathbb{C}/G_d$ of the coupling $\tau$, from which all other values of $\tau$ can be reached using the duality action. The fundamental domain we choose appears in Figure \ref{fig:FundamentalDomain}.
Note the existence of the special points in this domain - the point at the origin $\tau=0$ is the $XYZ$ model, while the point at $\tau=1$ is dual to $\tau=\infty$ and so describes (three copies of) the $D=3$ $\mathcal{N}=2$ Ising model. 
We will focus on the region $\mathcal{M}$ in the following.

\subsection{The lowest-dimension operator at fixed $R$-charge}

At the point $\tau=0$ on the conformal manifold $\mathcal{M}$, the theory has a complex one-dimensional moduli space of vacua.
This is equivalent to saying that the chiral ring relations will leave us with an infinite chiral ring, spanned by the set $\left\{X^n,\, Y^n,\, Z^n\right\}_{n\geqslant 0}$ \cite{Luty:1995sd}.
This means that the lowest operator dimension at large $R$-charge saturates the BPS bound at $\tau=0$,
\begin{equation}
    \Delta(Q_R)=\frac{D-1}{2}Q_R.
\end{equation}
This is consistent with the EFT analysis in \cite{Hellerman:2017veg}, in which the leading order EFT at large $R$-charge for the $XYZ$ model was the free Lagrangian.

The situation is very different when $0\neq\tau \in\mathcal{M}$.
The chiral ring relations leave us with a finite chiral ring, and so for sufficiently large $Q_R$ the lowest operator dimension will never saturate the BPS bound.
Moreover, by using the EFT analysis in \cite{Hellerman:2015nra}, we can even say that this failure of the saturation is parametric, \it i.e., \rm the lowest operator dimension scales as
\begin{eqnarray}
\Delta(Q_R)\propto Q_R^{3/2}\gg Q_R.
\end{eqnarray}
Note that the EFT analysis cannot predict the coefficient in front of $Q_R^{3/2}$.
Also note that this analysis was only valid when $Q_R$ is the largest parameter in the system.

In the following sections, we will study how the transition occurs between these two behaviours in the regime of large but not necessarily largest $Q_R$.
In other words, we will find a double-scaling parameter, which controls the transition for finite but large $Q_R$ and finite but small $\tau$. 
As a byproduct we will also compute the coefficient in front of $Q_R^{3/2}$ when $Q_R$ is the largest parameter in the system.

Another important question is determining the form of the lowest operator at large $R$-charge.
This can possibly be done using the $\epsilon$-expansion where the theory is weakly-coupled.
We do not carry out the computation in this paper, however.
We will discuss some properties of this operator in Section \ref{sec:what_operator}.

\section{Saddle point analysis at large $R$-charge}\label{sec:saddle_point}

We will now compute the classical configurations of the system which are solutions to the equations of motion under the fixed $R$-charge constraint on $S^{D-1}\times \mathbb{R}$.
We can then map the configuration with the lowest energy on the cylinder to the lowest-dimension operator at large $R$-charge, \it via \rm the state-operator correspondence.

\subsection{Known results in $D=4-\epsilon$ dimensions}

Since our WZ theory \eqref{eq:theory} is strongly coupled in the IR in three dimensions, we will study it in the $\epsilon$-expansion starting from four dimensions, where the interaction turns off in the IR and the theory flows to the Gaussian fixed-point for any value of $\tau$.

Most of the discussion from the previous section can be extended to $4-\epsilon$ dimensions, and so using the $\epsilon$-expansion in our analysis is justified. In particular, the conformal manifold exists for all intermediate dimensions between $3$ and $4$, and the theory has the same global symmetry discussed in the previous section. The duality group $G_d$ can also be extended to $4-\epsilon$ dimensions. 

The fixed point value of $\abs{g}^2$, which we call $\abs{g_{*}}^2$ (or simply $\abs{g}^2$ if there will be no confusion) hereafter, was computed to order $\epsilon^4$ in \cite{Ferreira:1996ug,Ferreira:1996az} using the $\epsilon$-expansion, as a function of $\tau$ and $\bar\tau$. We will only need the leading term, which is\footnote{We can use a $U(1)_R$ rotation to fix $g$ to be real, and so only its absolute value is relevant.}
\begin{equation}
\abs{g_{*}(\tau,\bar\tau)}^2=\frac{16\pi^2\epsilon}{3(2+\abs{\tau}^2)}+O(\epsilon^2)\;.
\label{eq:IRvalue}
\end{equation}
As expected, for small $\epsilon$ the theory becomes weakly coupled. Note that the duality $G_d$ demands that the combination $H(\tau,\bar\tau)\equiv\abs{g_{*}}^2(2+\abs{\tau}^2)$ be a duality invariant function, and this constrains the form of $\abs{g_*}^2$ to all orders in $\epsilon$.

\subsection{Saddle point configuration at large $R$-charge}

One can now compute the saddle point configuration which dominates at large $R$-charge, using the UV Lagrangian given in \eqref{eq:theory}. Concretely, we will look for the dominating saddle point for the theory, with chemical potential $\mu$ for the $R$-charge density turned on.
Working on the spherical spatial slice with unit radius, \it via \rm the state-operator correspondence, the energy associated with such a saddle point will be the lowest operator dimension at large $R$-charge.

For this particular case, the procedure of finding the saddle point configuration is equivalent to solving the equation of motion under the ansatz that (a) all bosonic fields are uniform in space and helical in time with frequency $\mu$,\footnote{
This is by no means a general procedure. For this model in particular, as the continuous global symmetry on the generic point of the moduli space is $U(1)$ and nothing else, there are no possible charge assignments that can take the lowest field configurations to be spatially inhomogeneous\cite{Alvarez-Gaume:2016vff,Hellerman:2017efx,Hellerman:2018sjf,Loukas:2017hic}.}
and that (b) all fermionic fields are turned off,\footnote{On the generic point of the conformal manifold, the VEV of the bosonic fields spontaneously brakes SUSY.
This determines the mass of the fermions as $m_{\phi}\propto d\Delta(Q)/dQ$ from the standard argument of massive Goldstini, which is parametrically large at large $Q$ \cite{Watanabe:2013uya,Hellerman:2015nra}.
This makes the ansatz to freeze all fermionic degrees of freedom consistent, as there will be no fermi sea involved.
One can also compute by hand the dispersion relation of fermions around the bosonic configurations to check the consistency of this ansatz, which will be conducted in Appendix \ref{sec:excitation}.
} along with the condition that the overall charge is kept fixed to $Q$. 

The most general helical solution under these assumptions is
\begin{equation}
\begin{pmatrix}
X\\
Y \\
Z 
\end{pmatrix}=
\begin{pmatrix}
A e^{i\mu t+\alpha}\\
B e^{i\mu t+\beta} \\
C e^{i\mu t+\gamma} 
\end{pmatrix}
\end{equation}
which is constrained by three equations of motion, one charge-fixing condition, and one freedom to eliminate the overall phase due to the $U(1)_R$ symmetry.
This means there is a two-dimensional space of solutions to the saddle-point equation.
Indeed, a general solution to the EOM can be parametrised as
\begin{equation}
\begin{pmatrix}
X\\
Y \\
Z 
\end{pmatrix}=\begin{pmatrix}
\frac{R}{\sin(\beta+\gamma+\arg\tau)} e^{i\mu t}\\
\frac{R}{\sin(-2\beta+\gamma+\arg\tau)} e^{i\mu t+i\beta} \\
\frac{R}{\sin(\beta-2\gamma+\arg\tau)} e^{i\mu t+i\gamma} 
\end{pmatrix}\;,
\label{eq:generic}
\end{equation}
where $R$ and $\mu$ are related by the EOM and determined by the fixed-charge condition, while $\beta$ and $\gamma$ are two free parameters as promised.

Now we need to find the saddle point with the lowest energy, which is the global minimum. 
Note that the EOM is symmetric under the combination of $t\mapsto -t$ and complex conjugation (plus possibly acting with the discrete global symmetry). 
To simplify the analysis, we will first assume that the global minimum is also symmetric under this combination, and then check numerically that this is indeed the case.
According to the ansatz, we now find that the lowest field configuration must be either one of the following four (after identifying configurations related by the discrete global symmetry $G$):
\begin{eqnarray}
\begin{pmatrix}
A e^{i\mu t}\\
0 \\
0 
\end{pmatrix},\quad
\begin{pmatrix}
A e^{i\mu t}/\sqrt{3}\\
A e^{i\mu t}/\sqrt{3} \\
A e^{i\mu t}/\sqrt{3} 
\end{pmatrix},\quad\begin{pmatrix}
A e^{i\mu t}/\sqrt{3}\\
A e^{i\mu t}/\sqrt{3} \\
A e^{i\mu t+2\pi i/3}/\sqrt{3} 
\end{pmatrix},\quad\begin{pmatrix}
A e^{i\mu t}/\sqrt{3}\\
A e^{i\mu t}/\sqrt{3} \\
A e^{i\mu t-2\pi i/3}/\sqrt{3} 
\end{pmatrix},
\label{eq:four-choices}
\end{eqnarray}
where $A$ and $\mu$ are determined by the equations of motion and the charge-fixing condition. 
Note that these are particular examples of \eqref{eq:generic}, \it via \rm a suitable rescaling.
The explicit form of $A$ and $\mu$ as a function of ${Q}$ will be given in \eqref{eq:mu} in the next subsection.

Now, the next question is which one of the four configurations \eqref{eq:four-choices} has the lowest energy, and so corresponds to the global minimum. For values of $\tau$ inside the fundamental domain of the duality group shown in Fig. \ref{fig:FundamentalDomain}, the answer turns out to be the first one.
This can be checked straightforwardly, or can be proven by comparing their energies at $\tau=0$ and at $\tau=\infty$ and using the Morse inequality, imitating the argument given in Section 2.2.2 of \cite{Baggio:2017mas}. To find the global minimum for other values of $\tau$, one can use the duality group $G_d$ to map $\tau$ to the new value. Note that this duality acts on the fields $X,Y,Z$ as well, and so it will take us to a different solution appearing in \eqref{eq:four-choices}. Thus for different regions in $\tau$-space, different saddle-point configurations dominate.

After finding the lowest configuration among the four candidates shown above, one can proceed to check if they are really locally the lowest ones among all configurations parametrised by $(\beta,\gamma)$. We performed this check numerically, and indeed there are no lower-energy solutions.
These pieces of information are enough to compute the lowest operator dimension at large $R$-charge for this model in the $\epsilon$-expansion.

\subsection{Truncation to $\phi^4$ theory}

We have found that we can restrict ourselves to the first configuration from \eqref{eq:four-choices}, since it is the global minimum at large $R$-charge inside the fundamental domain. This solution is particularly simple and simplifies the analysis immensely.
Remembering that the computation of the saddle point involves only the bosonic component $\phi_X$ of $X$, the Lagrangian can be truncated to
\begin{equation}\label{eq:truncated}
\mathcal{L}_{\rm simplified}\equiv
\abs{\partial\phi_X}^2+\frac{(D-2)^2}{4}\abs{\phi_X}^2+\frac{\abs{g_*\tau}^2}{4}\abs{\phi_X}^4\;,
\end{equation}
where $m$ is the conformal mass,
\begin{equation}
m=\frac{D-2}{2}.
\end{equation}
This Lagrangian coincides with the one used in \cite{Arias-Tamargo:2019xld,Badel:2019oxl,Watanabe:2019pdh} to compute the lowest operator dimension of the $O(2)$ Wilson-Fisher fixed point in $D=4-\epsilon$, with the identification
\begin{equation}\label{eq:dictionary}
{\abs{g_*(\tau,\bar\tau)}^2\abs{\tau}^2}
={\lambda_{*}^{\rm [BCMR]}},
\end{equation}
where $\lambda_{*}^{\rm [BCMR]}$ is $\lambda_{*}$ defined in \cite{Badel:2019oxl}.

By using the truncated Lagrangian \eqref{eq:truncated}, one can determine $A$ and $\mu$ for the first configuration in equation 
\eqref{eq:four-choices} as functions of $Q$.
Here we do not give the full expressions, but only give limiting behaviours at large or small $\abs{g\tau}^2Q$. 
$\mu$ can be determined as
\begin{eqnarray}
\mu&=&
\begin{cases}\displaystyle
m\left[1+\frac{\abs{g\tau}^2Q}{8m^3\alpha(D)}
-\frac{3\abs{g\tau}^4Q^2}{128m^6\alpha^2(D)}
+O\left(\left(\frac{\abs{g\tau}^2Q}{8m^3\alpha(D)}\right)^3\right)\right] & \displaystyle \left(Q\ll \frac{1}{{\abs{g\tau}^2}}\right)\\
\begin{split}
\displaystyle
\left(\frac{\abs{g\tau}^2Q}{4m^3\alpha(D)}\right)^{1/3}
\Bigg[1&+\frac{1}{3}\left(\frac{\abs{g\tau}^2Q}{4m^3\alpha(D)}\right)^{-2/3}
\\
&+O\left(\left(\frac{\abs{g\tau}^2Q}{4m^3\alpha(D)}\right)^{-4/3}\right)\Bigg]\end{split}
& \displaystyle \left(Q\gg \frac{16\pi^2}{{\abs{g\tau}^2}}\right)
\end{cases}\label{eq:mu}
\label{eq:A}
\end{eqnarray}
where $\alpha(D)$ is the area of the unit $D-1$-dimensional sphere
\begin{equation}
\alpha(D)=\frac{2\pi^{D/2}}{\Gamma(D/2)}\;.
\end{equation}
The parameter $A$ is given by
\begin{equation}
A^2=\frac{2(\mu^2-m^2)}{\abs{g\tau}^2}=\frac{2(\mu^2-\left(1-\epsilon/2\right)^2)}{\abs{g\tau}^2}
\label{eq:relationA2}
\end{equation}
in $D=4-\epsilon$ dimensions.

\section{Lowest operator dimension at large $R$-charge}\label{sec:dimension}

\subsection{General method}
\label{sec:method}

We follow the method presented in \cite{Badel:2019oxl} to compute the lowest operator dimension at fixed charge. The idea is to compute the energy of the lowest energy state at fixed charge on the sphere, which corresponds to the lowest operator dimension at fixed charge. This energy can be evaluated by using a saddle-point analysis at large $Q$ on the action deformed by a chemical potential term. The idea is to compute the correlator of the form
\begin{equation}
\Braket{\phi_Q|e^{-HT}|\phi_Q}
\end{equation}
for a generic arbitrary state with quantum number $Q$, and then take the limit $T\to \infty$ to extract the dimension of the lowest operator with charge $Q$, as long as the state $\Ket{\phi_Q}$ has an overlap with the lowest state.

Such a correlator can be evaluated using the path integral by plugging in the original action plus a chemical potential term.
We can then use the saddle-point analysis.
See Section 4.2 of \cite{Badel:2019oxl} for more details.

Therefore, as we have computed the saddle point configuration with lowest energy, all we need to do is to evaluate its energy \it via \rm the ordinary loop computations.
In the following we use the saddle-point of the form
\begin{eqnarray}
\begin{pmatrix}
A e^{i\mu t}\\
0 \\
0 
\end{pmatrix}
\label{eq:config}
\end{eqnarray}
which is the vacuum in the fundamental domain of the conformal manifold.
We will also discuss the effect of other saddle points in Section \ref{sec:validity}.
Our computations (especially the quantum corrections) will not be correct once we are out of the fundamental domain.

Because the coupling constant relating various fluctuations on top of such a saddle point is either $\abs{g\tau}^2$ or $\abs{g}^2$, these become the loop expansion parameter.
In other words, the structure of the operator dimension will look like, up to two-loop,
\begin{align}
\Delta(Q,\tau)=&\frac{1}{\abs{g}^2}\Delta_{0}(\lambda_1,\lambda_2)+\Delta_{1}(\lambda_1,\lambda_2)\\
&+\left(\abs{g}^2\Delta_{2,\abs{g}^2}(\lambda_1,\lambda_2)+\abs{g\tau}^2\Delta_{2,\abs{g\tau}^2}(\lambda_1,\lambda_2)\right)+\cdots
\end{align}
where $\Delta_{\rm cl}\equiv \frac{1}{\abs{g}^2}\Delta_{0}(\lambda_1,\lambda_2)$ is the classical piece, while $\lambda_1=\abs{g}^2Q$ and $\lambda_2=\abs{g\tau}^2Q$ are the two parameters we control in the double-scaling limit we will introduce later.

\subsection{Classical contribution}

Using \eqref{eq:dictionary} and the result presented in \cite{Badel:2019oxl}, we immediately find that the classical result $\Delta_{\rm cl}$ is:
\begin{equation}
\Delta_{\rm cl}=
\begin{cases}
\displaystyle
\textcolor{black}{
\begin{split}
mQ\Bigg[1&+\frac{1}{2}\left(\frac{\abs{g\tau}^2Q}{8m^3\alpha(D)}\right)
\\&- \frac{1}{2}\left(\frac{\abs{g\tau}^2Q}{8m^3\alpha(D)}\right)^2
+O\left(\left(\frac{\abs{g\tau}^2Q}{8m^3\alpha(D)}\right)^3\right)\Bigg]
\end{split}
}
& \textcolor{black}{
\displaystyle \left(Q\ll \frac{1}{{\abs{g\tau}^2}}\right)
}
\\\displaystyle
\textcolor{black}{
\begin{split}
\frac{3\alpha(D)}{\abs{g\tau}^2}\left(\frac{\abs{g\tau}^2Q}{4m^3\alpha(D)}\right)^{4/3}&
\Bigg[1+\frac{2}{3}\left(\frac{\abs{g\tau}^2Q}{4m^3\alpha(D)}\right)^{-2/3}
\\&+O\left(\left(\frac{\abs{g\tau}^2Q}{4m^3\alpha(D)}\right)^{-4/3}\right)\Bigg]
\end{split}
}
& 
\textcolor{black}{\displaystyle \left(Q\gg \frac{1}{{\abs{g\tau}^2}}\right)}
\end{cases}
\label{eq:classical}
\end{equation}
This makes it clear that there is a transition in the exponent of $\Delta$ as a function of $Q$ at around $Q\sim {1}/{\abs{g\tau}^2}$ --
When $Q\ll {1}/{\abs{g\tau}^2}$, the lowest operator dimension almost saturates the BPS bound, while for $Q\gg {1}/{\abs{g\tau}^2}$ the behaviour of $\Delta$ becomes semi-classical, in the sense that the exponent is simply a result of the classical dimensional analysis in four dimensions \cite{Hellerman:2015nra}. 

In the following, we will compute the one-loop correction to this expression. In particular, this will produce terms proportional to $Q^{4/3}\log Q$ when $Q\gg {1}/{\abs{g\tau}^2}$, slightly changing the exponent of $\Delta(Q)$ so that it will fit the classical dimensional analysis in $D=4-\epsilon$ dimensions.

\subsection{One-loop contribution}

One needs to evaluate the one-loop determinant around the lowest saddle point configuration. 
We conduct the computations separately for the different fields, and present them in Appendix \ref{sec:excitation}.
The computations will be done in three interesting regimes (as opposed to two as in the evaluation of $\Delta_{\rm cl}$).
This is because we have more than one small parameter in the system other than $1/Q$, as explained in the Introduction.

\subsubsection{Superfluid regime}
We start by studying the regime where $Q\gg 16\pi^2/(g\tau)^2$. 
In this limit the calculation from Appendix \ref{sec:excitation} gives
\begin{align}
\Delta_1&=\frac{\abs{\tau}^2+2}{8\abs{\tau}^2}\log\left(\frac{\abs{g\tau}^2Q}{4m^3\alpha(D)}\right)\left(\frac{\abs{g\tau}^2Q}{4m^3\alpha(D)}\right)^{4/3}+O\left(\left(\frac{\abs{g\tau}^2Q}{4m^3\alpha(D)}\right)^{4/3}\right)\nonumber\\
&=\textcolor{black}{\frac{2\pi^2\epsilon}{3\abs{g\tau}^2}\log\left(\frac{\abs{g\tau}^2Q}{4m^3\alpha(D)}\right)\left(\frac{\abs{g\tau}^2Q}{4m^3\alpha(D)}\right)^{4/3}+O\left(\left(\frac{\abs{g\tau}^2Q}{4m^3\alpha(D)}\right)^{4/3}\right)\;,\label{eq:Large_Q_1loop}
}
\end{align}
where we used \eqref{eq:IRvalue} in the second equality.

\subsubsection{Weak-coupling regime} 
Next we study the regime where $Q\ll 16\pi^2/\abs{g\tau}^2,\,16\pi^2/\abs{g}^2$. 
We find
\begin{align}
\Delta_1&=\left(1-\frac{1}{\abs{\tau}^2}+\sqrt{1-\frac{5}{\abs{\tau}^2}+\frac{4}{\abs{\tau}^4}}\right)\cdot\frac{\abs{g\tau}^2Q}{8m^3\alpha(D)}+O\left(\left(\frac{\abs{g\tau}^2Q}{16\pi^2}\right)^2\right)\nonumber\\
&=
\frac{\epsilon Q}{6}\left(1-\frac{4-2\sqrt{4-5\abs{\tau}^2+\abs{\tau}^4}-\abs{\tau}^2}{2+\abs{\tau}^2}\right)+
O\left(\left(\frac{\abs{g\tau}^2Q}{16\pi^2}\right)^2\right)\;.
\label{eq:small_Q_1loop}
\end{align}
We again used \eqref{eq:IRvalue} in the last equality.

\subsubsection{Near-BPS regime}

Finally, we study the regime where $16\pi^2/\abs{g}^2\ll Q\ll 16\pi^2/\abs{g\tau}^2$. 
The final expression is therefore only valid when $\abs{\tau}^2\ll 1$.
We also additionally require $\abs{\tau}^2$ to be the smallest parameter in the system, so that everything can be organised in terms of a small $\abs{\tau}^2$-expansion.
We have 
\begin{equation}
\begin{split}
\Delta_1&=k_0\left(\frac{\abs{g}^2Q}{8m^3\alpha(D)}\right)\\
&+\abs{\tau}^2\left[\frac{3}{2}\left(\frac{\abs{g}^2Q}{8m^3\alpha(D)}\right)^2\log \left(\frac{\abs{g}^2Q}{8m^3\alpha(D)}\right)+k_1\left(\frac{\abs{g}^2Q}{8m^3\alpha(D)}\right)\right]+O\left(\abs{\tau}^4\right),
\end{split}
\end{equation}
where $k_0(x)$ and $k_1(x)$ are undetermined functions with an asymptotic expansion
\begin{equation}
k_j(x)=\alpha_{j,4} x^2+ \alpha_{j,2} x+O(1).
\end{equation}
In the next section, the form of $k_0(x)$ will be determined on physical grounds.

\section{Final results and comparison with EFT}\label{sec:analysis}

By summing up the classical and the one-loop contribution to the lowest operator dimension at large $R$-charge, we are ready to get the final answer.
We will compare the result with the expectations from the effective field theory at large $R$-charge for large $Q$, and compare with the BPS bound near $\tau=0$.
We leave the comparison with Feynman diagram computations for future work.

\subsection{Semiclassical behavior for large $\epsilon\abs{\tau}^2 Q$}

We start by discussing the superfluid regime where $ Q\gg 16\pi^2/\abs{g\tau}^2$. When $Q$ is the largest parameter in the system, we expect the lowest operator dimension to follow the classical dimensional analysis \cite{Hellerman:2015nra}. We thus expect the behavior in this regime to follow that of the semiclassical phase.
Not only is this a general expectation from the EFT at large global charge,
this was also checked in the case of the Wilson-Fisher fixed point using the $\epsilon$-expansion. 

Therefore, as in the Wilson-Fisher case, we expect to find that the operator dimension in the $\epsilon$-expansion goes as
\begin{equation}
\left.\Delta(Q)\right|_{\text{leading in $Q$}}\propto Q^{\frac{4}{3}}\left (1+\frac{\epsilon}{9}\log Q\right)+O(\epsilon^2)\;,
\end{equation}
so that it is consistent with the expectation from the EFT, which is
\begin{equation}
\left.\Delta(Q)\right|_{\text{leading in $Q$}}= Q^{\frac{4-\epsilon}{3-\epsilon}}=Q^{\frac{4}{3}}\left (1+\frac{\epsilon}{9}\log Q\right)+O(\epsilon^2)\;.
\end{equation}
Indeed, the sum of classical and one-loop contributions from equations \eqref{eq:classical} and \eqref{eq:Large_Q_1loop} is
\begin{equation}
\Delta_{\textbf{cl}}+\Delta_1=\frac{3\alpha(D)}{\abs{g\tau}^2}
\left(\frac{\abs{g\tau}^2Q}{4m^3\alpha(D)}\right)^{4/3}
\left[1+\frac{\epsilon}{9}\log \left(\frac{\abs{g\tau}^2Q}{4m^3\alpha(D)}\right)\right]\;,
\end{equation}
so that the coefficient in front of the logarithm correctly reproduce the scaling $\frac{4-\epsilon}{3-\epsilon}$ up to order $O(\epsilon)$.

Moreover, since this can be understood as the EFT prediction for the resummation of loop diagrams to all-orders, we conclude that
\begin{align}
\Delta(Q)=\frac{3\alpha(D)}{\abs{g\tau}^2}
\left(\frac{\abs{g\tau}^2Q}{4m^3\alpha(D)}\right)^{\frac{D}{D-1}}
\left[1+O(\epsilon,\epsilon\abs{\tau}^2)+O\left(\left(\frac{\abs{g\tau}^2Q}{4m^3\alpha(D)}\right)^{-\frac{D-2}{D-1}}\right)\right].
\end{align}
In other words, we have computed the leading coefficient of the EFT at large $R$-charge for this model at leading order in $\epsilon$.

\subsection{Near-BPS behavior for small $\epsilon\abs{\tau}^2 Q$}

In this regime, the lowest operator is near BPS, in that its dimension is the BPS bound plus small corrections.
However, the correction behaves differently depending on whether $\epsilon Q$ is large or small.

\subsubsection{Small $\epsilon Q$ -- weak-coupling regime}

We discuss the regime in which $Q\ll 16\pi^2/\abs{g\tau}^2,\,16\pi^2/\abs{g}^2$. 
Because this regime includes the regime in which $\abs{\tau}^2\to 0$ while $\epsilon$ is also small, the result must reproduce the BPS bound at $\tau=0$, where there exists a moduli space of vacua and so the lowest operator at fixed $R$-charge is always BPS.

Indeed, summing up the classical and 1-loop results from equations \eqref{eq:classical} and \eqref{eq:small_Q_1loop}, we find
\begin{align}
\Delta(Q)=
\frac{D-1}{3}Q&+\frac{\epsilon Q}{6({2+\abs{\tau}^2})}\left(\abs{\tau}^2Q-4+2\sqrt{4-5\abs{\tau}^2+\abs{\tau}^4}+\abs{\tau}^2\right)\\&+O\left(\left(\frac{\abs{g\tau}^2Q}{16\pi^2}\right)^2\right)
\label{eq:small_Q_res}
\end{align}
which at $\tau=0$ reproduces the BPS bound exactly:
\begin{equation}
\Delta_{\rm BPS}=\frac{D-1}{2}R=\frac{D-1}{3}Q\;.
\end{equation}
Also note that the expression is always greater than the BPS bound for $Q\gg 1$.
We also find that the expression cannot be naively extrapolated to $Q\sim O(1)$ consistently without reservations unlike the $O(2)$ Wilson-Fisher case \cite{Arias-Tamargo:2019kfr,Badel:2019khk}, since the extrapolated operator dimension becomes below the BPS bound when $Q<Q_{\rm threshold}(\tau)$, where
\begin{align}
&Q_{\rm threshold}(\tau)= \frac{4-2\sqrt{4-5\abs{\tau}^2+\abs{\tau}^4}-\abs{\tau}^2}{\abs{\tau}^2}\\
&Q_{\rm threshold}(0)=\frac{3}{2}, \quad Q_{\rm threshold}(1)={3},
\end{align}
so that the extrapolation is clearly inconsistent.
This will be explained more in detail in Section \ref{sec:validity}.

We can also discuss the range of validity of our result \eqref{eq:small_Q_res} as a function of $\tau$. Note that the square root term may develop an imaginary part as we vary $\tau$, but this occurs only outside of the fundamental region shown in Figure \ref{fig:FundamentalDomain}. 
This is consistent with our analysis of the vacuum of the theory, which showed a first order phase transition to another vacuum as we exit the fundamental region, so that our analysis in not valid there. 
Instead, a different vacuum becomes the global minimum, which is consistent with the fact that the direct extrapolation develops an imaginary part.

\subsubsection{Large $\epsilon Q$, smallest $\abs{\tau}^2$ -- near-BPS regime}

Now we discuss the regime where $16\pi^2/\abs{g}^2\ll Q\ll 16\pi^2/\abs{g\tau}^2$. 
The final expression is therefore only valid when $\abs{\tau}^2\ll 1$.
We additionally require that $\abs{\tau}^2$ is the smallest parameter in the system, so that everything can be organised into the small $\abs{\tau}^2$-expansion.
As this region also includes the point where $\tau=0$, the BPS bound should be reproduced in this limit again.
This uniquely sets the form of $k_0(x)$ in the last section to be
\begin{equation}
k_0\left(\frac{\abs{g}^2Q}{8m^3\alpha(D)}\right)=\frac{\epsilon Q}{6}.
\end{equation}
Therefore, the final result for the operator dimension becomes
\begin{align}
\Delta(Q)=&
\frac{D-1}{3}Q+\frac{\abs{\tau}^2 Q}{2}\left(\frac{\abs{g}^2Q}{16\pi^2}\right)\left[1+\frac{3\abs{g}^2}{16\pi^2}\log \left(\frac{\abs{g}^2Q}{16\pi^2}\right)\right]\nonumber\\
&+O\left(\frac{\abs{g\tau}^2Q}{16\pi^2}\right)+O(\abs{\tau}^4)\;.
\end{align}
Note that this is not really a controlled expansion when $\epsilon\log Q\gg 1$, because of the term of the order $O\left((\epsilon\log Q)^k\right)$ inside the error term.
The precise validity of the expression is therefore when $\frac{1}{\epsilon}\ll Q\ll e^{1/\epsilon}$.
When $\epsilon\log Q\gg 1$, we can only speculate, for example, that they resum to an exponential or some other functions, which was exactly what happened in the superfluid regime.
It would be important to study the EFT of the theory at large charge directly in three dimensions in order to see what they resum to.

\section{Discussion of results}
\label{sec:validity}

We now discuss some final questions, including the range of validity of our answer and the form of the lowest-dimension operator. We will find that the operator whose dimension we computed was the one with $Q$-charge multiple of $3$.
We also find that there are corrections to our result of order $O(\epsilon e^{-Q})$ in the weak-coupling regime, so that our formula receives large non-perturbative corrections for $Q=O(1)$.
This is unlike the case of the $O(2)$ Wilson-Fisher fixed point and the $D=3$, $\mathcal{N}=2$ SUSY theory with superpotential $W=\Phi^3$, both treated in the $\epsilon$-expansion, where these corrections do not appear and the result in the weak-coupling regime is exactly valid even for $Q=O(1)$.
We will also explain the difference between these cases.

\subsection{What operator are we looking at?}\label{sec:what_operator}

The form of the operator we are looking at can in principle be determined.
As the degeneracy which existed at $\epsilon=0$ gets resolved for finite $\epsilon>0$, 
the theory selects some operators out of $\mathop{\rm Span} \left\{X^\alpha Y^\beta Z^\gamma|\alpha,\beta,\gamma\in\mathbb{N}_{\geqslant 0}\right\}$, whose dimensions one should be able to compute in the $\epsilon$-expansion in order to find the lowest one.
In this paper, we do not attempt to do this and leave it for future work.
However, we would like to point out that by going to the decoupling point $\tau=\infty$ on the conformal manifold, we are able to determine under which representation of the global symmetry group the lowest operator transforms.

At $\tau=\infty$, the theory is just three decoupled copies of the supersymmetric Ising model.
This means that the lowest operator is always of the form
\begin{align}
    X^{Q/3}Y^{Q/3}Z^{Q/3} \quad &\text{when} \quad Q\equiv 0 \quad (\text{mod } 3)\\
    X^{(Q-1)/3}Y^{(Q-1)/3}Z^{(Q-1)/3+1} \quad &\text{when} \quad Q\equiv 1 \quad (\text{mod } 3)\\
    X^{(Q-2)/3}Y^{(Q-2)/3+1}Z^{(Q-2)/3+1} \quad &\text{when} \quad Q\equiv 2 \quad (\text{mod } 3)
\end{align}
along with the permutations in $X$, $Y$, and $Z$.
This is true for any value of $Q$, and it doesn't require $Q\gg 1$ as long as we are in the weak-coupling regime, $\epsilon Q\ll 1$.
We will explain why this is the case in the next subsection.
It is in particular easy to see that these must be the form of the lowest operator at given $R$-charge in the semi-classical regime. -- The operator dimension becomes parametrically lower when the $R$-charge is distributed evenly among $X$, $Y$, and $Z$.
Since we do not have a phase transition as we go to the weak-coupling regime, this applies there as well.

As we have written down the form of the lowest operator at fixed $R$-charge, we can see what representation of the discrete global symmetry group it is in. The lowest operator at fixed $R$-charge is in the trivial representation $\mathbb{1}$ when $Q\equiv 0$, the $\mathbb{3}$ when $Q\equiv 1$, and the $\bar{\mathbb{3}}$ when $Q\equiv 2$, in the notation of \cite{Baggio:2017mas}.
One can now use the duality to map $\tau=\infty$ to $\tau=1$, and this is known to map $\mathbb{1}$, $\mathbb{3}$, or $\bar{\mathbb{3}}$ to itself, respectively \cite{Trautner:2016ezn}, so that the result is still the same at $\tau=1$.
As we do not expect any phase transitions on the fundamental domain of the conformal manifold, this should also carry over to the other values of $\tau$ on the fundamental domain.
We therefore conclude that the representation of the lowest operator at fixed $R$-charge of the global symmetry group falls into
\begin{align}
    \mathbb{1} \quad &\text{when} \quad Q\equiv 0 \quad (\text{mod } 3)\\
    \mathbb{3} \quad &\text{when} \quad Q\equiv 1 \quad (\text{mod } 3)\\
    \bar{\mathbb{3}} \quad &\text{when} \quad Q\equiv 2 \quad (\text{mod } 3)
\end{align}
for any value of $\tau$ and for any value of $Q$.

One can also compute their operator dimensions in the decoupling limit and compare with our result.
At $\tau=\infty$ and in the weak-coupling regime, they become, by using a similar analysis to the one above for the present model, 
\begin{align}
    \Delta_{\rm BPS}+\frac{\epsilon}{18}Q(Q-3)+O(\epsilon^2Q^2) \quad &\text{when} \quad Q\equiv 0 \quad (\text{mod } 3)\\
    \Delta_{\rm BPS}+\frac{\epsilon}{18}(Q-1)(Q-2)+O(\epsilon^2Q^2) \quad &\text{when} \quad Q\equiv 1 \quad (\text{mod } 3)\\
    \Delta_{\rm BPS}+\frac{\epsilon}{18}(Q-1)(Q-2)+O(\epsilon^2Q^2) \quad &\text{when} \quad Q\equiv 2 \quad (\text{mod } 3)
\end{align}
where as always, $\Delta_{\rm BPS}\equiv (D-1)Q/3$.
As we have explained in the previous subsection, note that this result is true for any value of $Q$, including $Q=O(1)$.
Now, we can compare this with \eqref{eq:small_Q_res} to notice that the expression we have there is only valid for $Q$, which is a multiple of $3$.
Although the entire argument was made in the weak-coupling regime, again because of the absence of the phase-transition, this also applies to the superfluid regime.
We can therefore conclude that the operator we are looking at is in the one-dimensional representation of the global symmetry group, for any value of $\tau$.
Note that this is not necessarily a contradiction since this only leads to subleading corrections in $Q$.

Also, by noting that the lowest operator at $Q=3$ is in the chiral ring and is of the form
\begin{equation}
    X^3+Y^3+Z^3-3\bar{\tau}XYZ+O(\epsilon^2),
\end{equation}
from the above reasoning we can speculate the form of the lowest operator at $Q=3n$ is
\begin{equation}
    \left(X^3+Y^3+Z^3-3\bar{\tau}XYZ+O(\epsilon^2)\right)^n.
\end{equation}
We do not check this statement using explicit computations in this paper, however.

Unfortunately, aside from the above reasoning, we still lack a clear understanding why the semi-classical configuration only describes the operator with $Q=3n$.
It would be interesting to have a systematic way of understanding the constraints on the lowest-dimension operator just by looking at the corresponding semi-classical configuration.\footnote{See the footnote $2$ of \cite{Hellerman:2018xpi} for a comment which might be related.}

\subsection{Does the result apply also for $Q=O(1)$?}

The result of the lowest operator dimension in the weak-coupling regime, \it i.e., \rm $1\ll Q\ll 1/\epsilon$, shown in \eqref{eq:small_Q_res}, is formally a perturbative expansion in $\epsilon$ and in $\epsilon Q$.
The consistency of the expansion, therefore, only requires $\epsilon\ll 1$ as well as $\epsilon Q\ll 1$, and does not require $Q\gg 1$.
So it seems like there is no obstruction in extrapolating $Q$ to be $O(1)$.
Indeed this was the case for the $O(2)$ Wilson-Fisher fixed-point in the $\epsilon$-expansion, where the operator dimension of $\phi^n$ computed in \cite{Badel:2019khk,Arias-Tamargo:2019kfr,Watanabe:2019pdh} \it via \rm the semi-classical method matched the Feynman diagram computation of the same quantity, which is also valid for $n=O(1)$.

Another case where the extrapolation is probably possible is the $D=3$, $\mathcal{N}=2$ SUSY theory with superpotential $W=\Phi^3$, treated in the $\epsilon$-expansion.
By doing the same analysis as in previous sections (or simply by noticing that $\tau=1$ point on our conformal manifold is nothing but three decoupled copies of this model), in the weak-coupling regime, the operator dimension of $\Phi^n$ becomes
\begin{eqnarray}
\Delta[\Phi^n]=\left(1-\frac{\epsilon}{3}\right)n+\frac{\epsilon}{2}n(n-1)+O(\epsilon^2n^2).
\end{eqnarray}
Note that the results for $n=1$ and for $n=2$ are compatible with supersymmetry and thus must be correct -- $\Delta[\Phi]$ satisfies the BPS bound, and we also have $\Delta[\Phi^2]=\Delta[\Phi]+1$, consistent with the fact that $\Phi^2$ is the descendent of $\Phi$.

On the other hand, our result for the cubic model in the weak-coupling regime cannot be consistently extrapolated to $Q=O(1)$.
For our cubic model, the lowest operator at $Q=3$ is in the chiral ring and its dimension is protected, but this is not reproduced in \eqref{eq:small_Q_res}, if we believe that the extrapolation is possible.

One possible reason for the mismatch is that there are non-perturbative corrections of them form $O(\epsilon e^{-Q})$ coming from the semi-classical computation. Concretely, we are computing the path-integral on $S^{D-1}\times\mathbb{R}$ with a certain boundary condition $B$ at $T=-\infty$ and at $T=\infty$ for the field configurations,
\begin{eqnarray}
\int_{B} \mathcal{D}X_i\, e^{-S_{\mathrm{eff}}}
\end{eqnarray}
where $S_{\mathrm{eff}}$ is the action with chemical potential added.
The boundary condition $B$ in this case is, as explained in 
\eqref{eq:config},
\begin{eqnarray}
\begin{pmatrix}
X(t)\\
Y(t) \\
Z(t) 
\end{pmatrix}
=
\begin{pmatrix}
A e^{i\mu t}\\
0 \\
0 
\end{pmatrix}
 \quad \text{for} \quad t\to\pm\infty
\end{eqnarray}

Now, what we did in Section \ref{sec:dimension} is to take a saddle-point configuration which is simply helical in time,
but there are other configurations.
As is shown in \eqref{eq:generic}, there are multiple solutions to the equation of motion of this system with chemical potential added.
Because the space of such solutions is topologically nontrivial, there are instanton-like solutions to the EOM, and they also contribute to the lowest dimension of the operator at fixed $R$-charge.
Therefore we conclude that there are instanton-like corrections to the formula 
\eqref{eq:small_Q_res}, which are of order $O(\epsilon e^{-Q})$ (since the action of such instantons would scale as $Q$, and because there can be no corrections to the formula at order $O(\epsilon^0)$, \it i.e., \rm in $D=4$). 
In addition, at the boundary of the fundamental region there are additional vacua, and we expect instanton-like contributions from there as well.

We conclude that we cannot extrapolate \eqref{eq:small_Q_res} to small $Q$ because of instanton-like contributions.
This seems to imply that in the case where one has only one solution to the EOM, such as the $O(2)$ Wilson-Fisher fixed-point, the semi-classical result from the one and the only saddle-point reproduces the Feynman diagram computation even when the global charge is not necessarily large, as there are no instanton-like corrections for small charge.

\section{Conclusion and Outlook}\label{sec:conclusion}

In this paper we computed the lowest operator dimension at large $R$-charge of the $D=3$, $\mathcal{N}=2$ supersymmetric theories with superpotential 
\begin{equation}
    W=g\left(XYZ+\frac{\tau}{6}\left(X^3+Y^3+Z^3\right)\right)
\end{equation} 
using the $\epsilon$-expansion.
The computation was conducted in three different regimes where it simplifies, 
\it i.e., \rm where the large parameter $Q_R$ is much larger than, in between, or much smaller than $\frac{1}{\epsilon}$ and $\frac{1}{\epsilon\abs{\tau}^2}$. 
Note that the first regime would have been inaccessible by simply using the supersymmetry algebra as the operator we are interested in is usually way above the BPS bound.
Note also that the first and second regimes cannot be easily be accessed by conventional Feynman diagrams as the loop contributions are enhanced by a large number of insertions, proportional to $Q_R$.

The actual computation followed a recently proposed technique of combining the $\epsilon$-expansion with the large-charge EFT.
Among other things, we find (1) that the lowest operator dimension scales as $Q_R^{\frac{4-\epsilon}{3-\epsilon}}$ when the $R$-charge is much larger than $\frac{1}{\epsilon\abs{\tau}^2}$, and we calculated its precise coefficient to leading order in the $\epsilon$-expansion, and (2) that it goes as the BPS bound plus a small correction when $\epsilon\abs{\tau}^2$ is much smaller than the inverse of the $R$-charge.
This means that there is a double-scaling parameter $\epsilon\abs{\tau}^2Q_R$ in the theory that interpolates between the superfluid phase and the near-BPS phase at large global charge.
Although this is another instance of a case demonstrated in \cite{Arias-Tamargo:2019kfr,Badel:2019khk,Watanabe:2019pdh} in this regard,
what is new about this example is that the entire transition can be expected to happen in full-fledged unitary theories in three dimensions, since we can consistently extrapolate $\epsilon\to 1$ later on and still observe the transition as a function of the exactly marginal operator $\tau$.
In the $O(2)$ Wilson-Fisher theory at large global charge studied in \cite{Arias-Tamargo:2019kfr,Badel:2019khk,Watanabe:2019pdh}, the double-scaling parameter was $\epsilon Q$, so that the transition was happening in fractional spacetime dimensions.

Another new aspect of the present case was the existence of a duality.
Due to the presence of the duality, 
we could see that there is a first-order phase transition of the lowest operator dimension at large charge on the $\tau$-plane -- there are four operators mapped to one another by the duality action, which are exchanged once we exit the fundamental domain of the conformal manifold $\mathcal{M}$.
In addition, combining the duality with the existence of a decoupling limit on the conformal manifold, we were able to see that the operator we are looking at should always have a $Q$-charge ($3/2$ times the $R$-charge) which is a multiple of $3$.
For other operators which do not satisfy the condition, we cannot compute the operator dimension precisely, although the correction to their dimension is subleading in $1/Q$.

Having multiple solutions to the EOM at fixed global charge was also a new aspect of this model compared to the Wilson-Fisher case.
This caused the result in the weak-coupling regime ($\epsilon\abs{\tau}^2Q_R\ll 1$) to be supplemented by nonperturbative corrections of order $O(\epsilon e^{-Q_R})$, whereas this type of contribution was nonexistent for the Wilson-Fisher case.

We list several interesting future  directions.
First, there are a few generalizations of this calculation that can be done. One such generalization is the conformal manifold of $\mathcal{N}=4$ $SU(N)$ SYM. In terms of $\mathcal{N}=1$ superfields, this theory has a vector multiplet and three adjoint chiral multiplets $X,Y,Z$ with a superpotential term $W\propto \tr X[Y,Z]$. For $N>2$, this theory has a complex three-dimensional conformal manifold which preserves $\mathcal{N}=1$ SUSY. Focusing on a single direction corresponding to deforming the superpotential by $\tr(X^3+Y^3+Z^2)$, we find that this theory looks similar to our WZ model, and one can consider the theory at large $R$-charge. 

Second, it would be interesting to write down the EFT of the present theory at large $R$-charge, which is applicable to both at $\tau=0$ and elsewhere on the conformal manifold.
This will allow us to study the theory directly in three dimensions, without resorting to the $\epsilon$-expansion.
We would have to use the global symmetry and the duality fully in order to write down all possible operators inside the EFT.
We might also have to use the fact that the theory becomes three decoupled copies of the SUSY Ising model at $\tau=1$.

In addition, studying the causality constraint of the theory at large $R$-charge might be interesting.
At $\tau=0$, there is already one known causality constraint on the subleading terms in the  EFT at large $R$-charge, which is visible at the level of the first non-protected operator at large and fixed $R$-charge in terms of the operator dimension \cite{Hellerman:2017veg}.
As there is no such thing known for the superfluid EFT,
it will be quite intriguing if we can find a similar constraint on the EFT when $\tau\neq 0$.
We may be able to start by computing the dimension of the low-lying spectrum above the lowest dimension operator at large $R$-charge, although such a computation is expected to be tedious.
Because the causality constraint at $\tau=0$ means that a certain coefficient of such operator dimensions can only be negative, we would be able to observe how it changes as one moves out of $\tau=0$.
This may also be interesting in the context of the unified EFT at large charge in the last paragraph.

It is also interesting to reproduce some of the results using Feynman diagram computations.
For the case of the $O(2)$ Wilson-Fisher theory, it was pointed out that the lowest operator dimension computed for small $\epsilon Q$ but large $Q$ matched the Feynman diagram computation of the dimension of the operator $\phi^Q$, even for small $Q$ \cite{Arias-Tamargo:2019kfr, Badel:2019khk}.
However, one should be careful in comparing our result with the Feynman diagram computation, because the result we have in this paper only applies to $Q$-charge being the multiple of $3$, as well as because of the non-perturbative correction we have of order $O(\epsilon e^{-Q})$.
Nevertheless it would be still interesting to see how much can be learned about the low-$Q$ spectrum from such an analysis.
Also, it is very interesting to understand why the method in this paper only applied to the case where the $Q$-charge is a multiple of $3$.
See footnote $2$ of \cite{Hellerman:2018xpi} for a small comment which might be related.

Finally, it would be interesting to expand our knowledge about various phases at large global charge and the transitions among them.\footnote{See \cite{Delacretaz:2020nit} for another approach using hydrodynamics. See also \cite{Loukas:2018zjh,delaFuente:2020yua} for a similar question in the dual gravity theory.}
We can for example consider theories where at large $R$-charge the VEV breaks only some of the supersymmetry spontaneously. In our case, we studied the transition between phases where the $R$-symmetry and the supersymmetry are unbroken ($\tau=0$) to  completely broken ($\tau\neq 0$), and so the existence of SUSY was not useful for our computations. If we study transitions where only part of the supersymmetry breaks, the computations should naively be much simpler.
In particular, although the lowest dimension operator at fixed $R$-charge of such theories presumably always saturates the BPS bound, it might be interesting to study OPE coefficients among them, or the low-lying spectrum above them.

Another example of an interesting phase at large charge involves the Fermi sea.
This can be easily realised in a system of a free massless Dirac fermion in three dimensions, and the EFT on top of the Fermi sea will be built out of the particle-hole excitations and will be gapless.
It would be interesting to write down this EFT in the context of the EFT at large charge, for one thing \cite{Polchinski:1992ed}.
We might also be able to find a transition between the phase including the Fermi sea and the phase without the Fermi sea, possibly by using the boson-fermion dualities in three dimensions.

\section*{Acknowledgements}
\begin{sloppypar}
The authors are grateful to Ofer Aharony, Micha Berkooz, Shai Chester, Simeon Hellerman, Zohar Komargodski, Ohad Mamroud, Domenico Orlando, Himanshu Raj, Shlomo Razamat, Susanne Reffert, Tal Sheaffer and Yifan Wang for valuable discussions.
The work of MW is supported by
the Foreign Postdoctoral Fellowship Program of the Israel Academy of Sciences and Humanities.
This work was supported by an Israel Science Foundation center for excellence grant (grant number 1989/14) and by the Minerva foundation with funding from the Federal German Ministry for Education and Research.

\end{sloppypar}

\appendix

\section{Computation of the one-loop determinants}
\label{sec:excitation}

Following \cite{Badel:2019oxl}, we compute the excitation spectrum on top of the saddle point for the various fields of the theory, and then compute the one-loop determinant. The divergences will be regularised by using $\zeta$-function regularisation. This is known to be equivalent to the $\epsilon$-expansion at least up to one-loop, which we already used to find the value of the coupling constant at the IR fixed-point.
For more information about how to consistently renormalise the one-loop determinant in a similar case, see \cite{Baggio:2017mas}.

\subsection{Excitation spectrum and the one-loop determinants}

In order to evaluate the $1$-loop determinant around the saddle point of interest above, one first needs to know the excitation spectrum on top of such a vacuum.
This can be done by writing the Lagrangian and expanding it up to quadratic order in fluctuations. Since we are studying the system on a spherical spatial slice, the dispersion relation $\omega$ will depend on the eigenvalues of the Laplacian or the Dirac operator on $S^{D-1}$. These are respectively given as
\begin{equation}
p_B(\ell)=\sqrt{\ell(\ell+D-2)},
\end{equation}
and
\begin{equation}
{p}_F(\ell)=\pm\left(\ell+\frac{D-1}{2}\right),
\end{equation}
where $\ell=0,\,1\,\cdots$ is the standard label of eigenvalues of those operators on the sphere. These eigenvalues are degenerate with multiplicity
\begin{equation}
n_B(\ell)=\frac{(2\ell+D-2)\Gamma(\ell+D-2)}{\Gamma(D-1)\Gamma(\ell+1)}
\end{equation}
and
\begin{equation}
n_F(\ell)=2\frac{\Gamma(\ell+D-1)}{\Gamma(D-1)\Gamma(\ell+1)}
\end{equation}
respectively \cite{Camporesi:1995fb}.\footnote{
To be precise, the factor of two in the expression for $n_F(\ell)$ should be the dimension of the smallest spinor representation in $D$ dimensions, which is $2^{D/2-1}$ for even $D$ and $2^{(D-1)/2}$ for odd $D$.
However, as we are in $4-\epsilon$ dimensions and we would like to keep the dimension of the fermion representation intact, we will stick to this expression hereafter.
In other words, we will start from $\mathcal{N}=1$ theory in four dimensions with two two-component spinors, to reach the $\mathcal{N}=2$ theory in three dimensions in question -- left/right-handed fermions will simply become different fermions in three dimensions, corresponding to two different SUSY charges.}

The excitation spectrum can then be used to infer the value of the one-loop determinants $\Delta_1$, which can be done in the standard method of computing the Casimir energy.
This just means summing over the zero-point energies for all modes on the spatial slice \cite{Badel:2019oxl}. Concretely, for bosons with dispersion relation $\omega_B(\ell)$, the one-loop determinant is
\begin{equation}
\Delta_{1,B}=\frac{1}{2}\sum_{\ell=0}^{\infty}n_B(\ell)\omega_B(\ell)\;.
\end{equation}
Similarly, for fermions with dispersion relation $\omega_F(\ell)$, the one-loop determinant is given by
\begin{equation}
\Delta_{1,F}=-\frac{1}{2}\sum_{\ell=0}^{\infty}n_F(\ell)\omega_F(\ell).
\end{equation}
Note the minus sign due to the fermionic statistics.

\subsubsection{Contribution from $\phi_X$}

The excitation spectrum of the bosonic component of $X$ has been already worked out in \cite{Badel:2019oxl}. They find
\begin{equation}
\left(\omega^{\phi_X}_{\pm}(\ell)\right)^2
=p_B(\ell)^2+3\mu^2-m^2\pm \sqrt{(3\mu^2-m^2)^2+4p_B(\ell)^2\mu^2}\;,
\end{equation}
where $\omega^{\phi_X}_{+}$ and $\omega^{\phi_X}_{-}$ are the excitation spectrum for the radial and the angular mode of $\phi_X$, respectively.
By using this, the one-loop determinant $\Delta_0^{\phi_X}$ from the bosonic component of $X$ can be written as
\begin{equation}
\Delta_1^{\phi_X}\equiv\frac{1}{2}\sum_{\ell}n_b(\ell)\left(\omega_+^{\phi_X}(\ell)+\omega_-^{\phi_X}(\ell)\right)\equiv \sum_{\ell} f_{\phi_X}(\ell).
\end{equation}

\subsubsection{Contribution from $\psi_X$}

The excitation spectrum of the fermionic component of $X$ can be read off from the Lagrangian truncated up to quadratic orders in fluctuations, which is 
\begin{equation}
\mathcal{L}_{\rm quad}=i\left(\partial_{\mu}\bar\psi_X\right)\bar\sigma^\mu\psi_X-\frac{{g}{\tau}}{2}\langle\phi_X\rangle\psi_{X}\psi_{X}-\frac{{\bar{g}}{\bar\tau}}{2}\langle\bar\phi_X\rangle\bar\psi_{X}\bar\psi_{X},
\end{equation}
where $\langle\phi_X\rangle=A e^{i\mu t}$.
As mentioned in the beginning of this subsection, we wrote the above Lagrangian down assuming the theory has $\mathcal{N}=1$ supersymmetry in four dimensions, with the two-component notation for the fermion.
The notation for spinor indices is inherited from the one given in \cite{Wess:1992cp}.
This is convenient since it can have a natural interpretation in terms of the original $D=3$, $\mathcal{N}=2$ theory.

Following \cite{Hellerman:2015nra}, we rewrite this Lagrangian by redefining the fermionic fields as follows,
\begin{equation}
\Psi=\psi_Xe^{i\mu t/2},\quad \bar\Psi=\bar\psi_Xe^{-i\mu t/2},
\end{equation}
which leads to
\begin{equation}
\mathcal{L}_{\rm quad}=
i\left(\partial_\mu \bar\Psi\right)\bar\sigma^\mu\Psi
-\frac{\mu}{2}\bar\Psi\bar\sigma^0\Psi
-\frac{{g}{\tau A}}{2}\Psi\Psi
-\frac{{\bar{g}}{\bar\tau A}}{2}\bar\Psi\bar\Psi.
\end{equation}
The excitation spectrum computed from this Lagrangian can be obtained by diagonalising the mass matrix, which leads to
\begin{align}
	\omega^{\psi_X}_{\pm}(\ell)&=
	\sqrt{\left(\abs{p_F(\ell)}\pm\frac{\mu}{2}\right)^2+\abs{g\tau}^2A^2}\\
	&=\sqrt{\left(\abs{p_F(\ell)}\pm\frac{\mu}{2}\right)^2+2(\mu^2-m^2)}
	.
\end{align}
This explicitly checks that the fermion is gapped for $\tau\neq 0$, while for $\tau=0$ the configuration we consider is simply supersymmetric, which means for both cases it was consistent to turn off the $X$ fermion when computing the dominating saddle point configuration.

The one-loop determinant $\Delta_0^{\psi_X}$ from the fermionic component of $X$ can be written as
\begin{equation}
\Delta_1^{\psi_X}\equiv-\frac{1}{2}\sum_{\ell}n_f(\ell)\left(\omega_+^{\psi_X}(\ell)+\omega_-^{\psi_X}(\ell)\right)\equiv \sum_{\ell} f_{\psi_X}(\ell).
\end{equation}

\subsubsection{Contribution from $\phi_{Y,Z}$}

The excitation spectrum of the bosonic components of $Y$ and $Z$ can be read out from the Lagrangian truncated up to quadratic orders in fluctuation, which is 
\begin{equation}
\begin{split}
\mathcal{L}_{\rm quad}=
\abs{\partial\phi_Y}^2+\abs{\partial\phi_Z}^2
&+\abs{g}^2\abs{\Braket{\phi_X}}^2\left(\abs{\phi_Y}^2+\abs{\phi_Z}^2\right)
\\
&+\frac{\abs{g}^2}{2}\left(\tau\Braket{\phi_X}^2\bar\phi_Y\bar\phi_Z+\bar\tau\Braket{\bar\phi_X}^2\phi_Y\phi_Z\right)
\end{split}
\end{equation}
where $\langle\phi_X\rangle=A e^{i\mu t}$.

Again it helps to rewrite this Lagrangian by redefining the fields as follows (note that the unusual form of the redefinition relating a quantity with a bar to one without bars),
\begin{equation}
Y=e^{i\mu t}\bar\phi_Y,\quad Z=e^{-i\mu t}\phi_Z,
\end{equation}
which leads to
\begin{equation}
\begin{split}
\mathcal{L}_{\rm quad}
=
\abs{D Y}^2+\abs{D Z}^2+(\abs{g}^2A^2&+m^2)(\abs{Y}^2+\abs{Z}^2)
\\&+
\frac{\abs{g}^2A^2}{2}\left(\tau Y\bar Z+\bar\tau \bar Y Z\right),
\end{split}
\end{equation}
where 
\begin{equation}
D_0Y=(\partial_0-i\mu)Y,\quad D_0Z=(\partial_0+i\mu)Z,\quad D_i=\partial_i.
\end{equation}

The dispersion relation can be read off from this Lagrangian as a solution to the following quartic equation,
\begin{equation}
\begin{split}
\left|
\begin{matrix}
-(\omega+\mu)^2+p_B(\ell)^2+m^2+\abs{g}^2A^2 & A^2\abs{g}^2\bar\tau /2\\
A^2 \abs{g}^2\tau/2 & -(\omega-\mu)^2+p_B(\ell)^2+m^2+\abs{g}^2A^2
\end{matrix}
\right|
=0,
\end{split}
\end{equation}
where the bracket means the determinant of the matrix inside.
The final result for the dispersion relation is therefore
\begin{align}
	\left(\omega^{\phi_Y,\phi_Z}_{\pm}\right)^2&=
	p_b(\ell)^2+\mu^2+m^2+\abs{g}^2A^2
	\\
	&\qquad\qquad \pm\sqrt{4\mu^2\left(p_B(\ell)^2+\abs{g}^2A^2+m^2\right)+\left(\frac{\abs{g}^2\abs{\tau}A^2}{2}\right)^2}
	\\
	&
	=\mu^2\left(1+\frac{2}{\abs{\tau}^2}\right)+m^2\left(1-\frac{2}{\abs{\tau}^2}\right)+p_B(\ell)^2\\
	&\qquad\qquad\pm\sqrt{
		4\mu^2\left(p_B(\ell)^2+m^2\right)+\frac{(\mu^2-m^2)(9\mu^2-m^2)}{\abs{\tau}^2}
	}
\end{align}
This expression coincides with the one for $\phi_X$ at $\tau=1$, as expected from the enhanced $U(1)^3$ global symmetry there.

Finally, the one-loop determinant $\Delta_0^{\psi_X}$ from the bosonic component of $Y$ and $Z$ can be written as
\begin{equation}
\Delta_1^{\phi_{Y,Z}}\equiv-\frac{1}{2}\sum_{\ell}n_f(\ell)\left(\omega_+^{\phi_{Y,Z}}(\ell)+\omega_-^{\phi_{Y,Z}}(\ell)\right)\equiv \sum_{\ell} f_{\phi_{Y,Z}}(\ell).
\end{equation}

\subsubsection{Contribution from $\psi_{Y,Z}$}

The excitation spectra of the fermionic component of $Y$ and $Z$ can be read off from the Lagrangian truncated up to quadratic orders in fluctuation, which reads,
\begin{equation}
\begin{split}
\mathcal{L}_{\rm quad}=i\left(\partial_{\mu}\bar\psi_Y\right)\bar\sigma^\mu\psi_Y+i\left(\partial_{\mu}\bar\psi_Z\right)\bar\sigma^\mu\psi_Z-{g}\langle\phi_X\rangle\psi_{Y}\psi_{Z}-{\bar{g}}\langle\bar\phi_X\rangle\bar\psi_{Y}\bar\psi_{Z}
\end{split}
\end{equation}
where again $\langle\phi_X\rangle=Ae^{i\mu t}$.

Employing similar field redefinitions as the previous ones, the excitation spectrum becomes
\begin{align}
	\omega^{\psi_Y,\psi_Z}_{\pm}(\ell)&=
	\sqrt{\left(\abs{p_F(\ell)}\pm\frac{\mu}{2}\right)^2+\abs{g}^2A^2}\\
	&=\sqrt{\left(\abs{p_F(\ell)}\pm\frac{\mu}{2}\right)^2+\frac{2(\mu^2-m^2)}{\abs{\tau}^2}}
	.
\end{align}
This expression again coincides with the one for $\psi_X$ at $\tau=1$, which is also expected from the enhanced $U(1)^3$ global symmetry there.

Finally, the one-loop determinant $\Delta_0^{\psi_{Y,Z}}$ from the fermionic component of $Y$ and $Z$ can be written as
\begin{equation}
\Delta_1^{\psi_{Y,Z}}\equiv-\frac{1}{2}\sum_{\ell}n_F(\ell)\left(\omega_+^{\psi_{Y,Z}}(\ell)+\omega_-^{\psi_{Y,Z}}(\ell)\right).
\end{equation}

\subsection{Regularisation of the one-loop determinants}

Of course, the infinite sums presented above are divergent, and we need to regularise and renormalise them.
Here, following \cite{Badel:2019oxl}, we will use $\zeta$-function regularisation.
Again this is known to be equivalent to the $\epsilon$-expansion at least up to one-loop -- to be precise, one should use the method employed in \cite{Baggio:2017mas}, where the computation is done entirely in $\zeta$-function regularisation, before computing the value of the coupling constant on the IR fixed-point, by demanding that the final result be independent of the cut-off.
However, we will stick to the prescription employed in \cite{Badel:2019oxl}, which we review in the following.

We now compute these regularised sums. All of our sums are of the form
\begin{equation}\label{eq:sum}
\Delta_1=\sum_\ell F(\ell,\mu)
\end{equation}
where
\begin{equation}
F(\ell,\mu)=\sum_{k=1}^{\infty}{c_k(\mu)}{\ell^{D-k}}\;.
\end{equation}
We will focus on calculating the sum separately in the three regimes discussed in the Introduction. First, we will discuss the large-$Q$ regime, were $B=\mu\gg1$. Next, we calculate the sum for small $Q$, where $y=\mu-m\ll 1$. Finally, we will compute it for intermediate $Q$ where $y=\mu-m\ll 1$ and also $B=\frac{\sqrt y}{\abs{\tau}}\gg1$. The sum simplifies in these limits, and they are precisely the limits which concern us.

Note that the sums over the first $D+1$ terms $c_1,...,c_{D+1}$ diverge, and so they must be regulated. In the $\epsilon$-expansion, we can expand each $c_k$:
\begin{equation}
c_k=c_{k,0}+c_{k,1}\epsilon+O(\epsilon^2)\;.
\end{equation}

\subsubsection{Calculating the sum when a parameter $B$ is large}

We compute the sum \eqref{eq:sum} when there is a large parameter $B$ (this $B$ will be taken to be either $\mu$ or $\frac{\sqrt y}{\abs{\tau}}$ in the following). It will be important that $F(\ell,B)$ has a polynomial dependence on $\ell$ or on $B$ at large $\ell$ or large $B$, respectively.
Although what we mean by a ``large parameter'' here is not specified completely (when $\mu$ is large, $B^2$ for example is large), we will fix $B$ by demanding that $F(\ell,B)$ has the polynomial behaviour with the same order at large $\ell$ and at large $B$, \it i.e., \rm
\begin{equation}
F(\ell,B)\to
\begin{cases}
 B^p  &(B\to \infty)\\
 \ell^p &(\ell\to \infty)
\end{cases}
\end{equation}

The Euler-Maclaurin formula can be used to approximate this sum, by introducing a new large parameter $\tilde{B}\equiv\alpha B$ as a cut-off ($\alpha$ is some $O(1)$ constant),
\begin{equation}
\sum_\ell F(\ell,B)=\sum_{\ell=0}^{\tilde{B}} F(\ell,B)+\int^{\infty}_{\tilde{B}} d\ell\, F(\ell,B)+G(B),
\end{equation}
where $G$ can be written using the Bernoulli numbers $B_n$ as
\begin{equation}
G(B)=-F(\ell=\alpha B,B)-\sum_{k=1}^{\infty}\frac{B_{2k}}{\Gamma(2k+1)}\left.\frac{d^k}{d\ell^k}\right|_{\ell=\alpha B}F(\ell,B).
\end{equation}
Now, the integral can be evaluated (in dimensional regularisation) as
\begin{equation}
\int^{\infty}_{\tilde{B}} d\ell\, F(\ell,B)=\sum_{k=1}^{\infty}\frac{c_k}{k-(D+1)}\left(\alpha B\right)^{D+1-k}.
\end{equation}
This can also be expressed in the $\epsilon$-expansion as follows,
\begin{equation}
\int^{\infty}_{\tilde{B}} d\ell\, F(\ell,B)=\frac{c_{5,0}}{\epsilon}-c_{5,0}\log B + H(B) +O(\epsilon),
\end{equation}
in which $H(B)$ is a polynomial in $B$.

All in all, the sum therefore becomes, after regularisation,
\begin{equation}
\sum_\ell F(\ell,B)=-c_{5,0}\log B+\sum_{\ell=0}^{B} F(\ell,B)+ G(B)+H(B)\equiv -c_{5,0}\log B+K(B).
\end{equation}
Here, $K(B)$ is expected to be a polynomial in $B$, whose precise form depends on $c_k(B)$.

\subsubsection{Calculating the sum when no parameter is large}

We now discuss the computation of the sum \eqref{eq:sum} when no parameter is large. This time, we regularise the divergent parts of the sum using $\zeta$-functions. Explicitly, this leads to
\begin{align}
\sum_\ell F(\ell,B)
&=
\sum_{k=1}^{5}c_{k}\zeta(k-D)+F(0)+\sum_{\ell=1}^{\infty}\tilde{F}(\ell)\\
&=
\frac{c_{5,0}}{\epsilon}+c_{5,1}+\gamma c_{5,0}+\sum_{k=1}^{4}c_{k,0}\zeta(k-4)+F(0)+\sum_{\ell=1}^{\infty}\tilde{F}(\ell)+O(\epsilon)\label{eq:small_Q_dimension}
\end{align}
where $\tilde{F}(\ell)$ is a convergent sum,
\begin{equation}
\tilde{F}(\ell)=\sum_{k=6}^{\infty}{c_k(\mu)}{\ell^{D-k}}.
\end{equation}
Also $\gamma$ here is the Euler gamma.
For small $y=\mu-m$, one can now expand the expressions in $y$ and compute them order by order.

\subsubsection{A comment about the regularization scheme}

Here, we have no reason why this regularization scheme should match the dimensional regularization and the $\overline{\mathrm{MS}}$ scheme which was used to compute the value of the coupling constant on the IR fixed-point.
The difference in those schemes, however, is not much of an issue here, because of the universality of the $O(\epsilon^0)$ piece of the $\beta$-function at $1$-loop.
In other words, possible scheme difference would only multiply some constant factor to the value of the coupling constant in the IR.

Now, in the superfluid regime, such a multiplication would be inconsistent, because then that would fail to reproduce the EFT prediction that the operator dimension must scale as $Q^{(4-\epsilon)/(3-\epsilon)}$.
Note that the contribution at $1$-loop which goes as $Q^{4/3}\log Q$ must come with a precise coefficient, relative to the $Q^{4/3}$ term in the $0$-loop contribution.
This means that, at least at $1$-loop, the result presented below using the above regularization is consistent in the superfluid regime.

Likewise, in the weak-coupling region, you can argue that the regularization above should reproduce the correct result.
This is rather subtle, but the argument goes as follows --
One can imagine a different scheme in which we evaluate the classical saddle point using the Lagrangian, with conformal mass $m^2=(1-\frac{\epsilon}{2})^2$ replaced with $m^2|_{\epsilon=0}=1$.
This is actually the regularization used in \cite{Badel:2019khk}, and then in that case, the term $-\frac{\epsilon Q}{2}$ in the $0$-loop result in our paper will be absorbed into the $1$-loop determinant, but as a function which does not depend on $\epsilon$, but on $\abs{g\tau}^2 Q$.
The result we get should be scheme-independent, and replacing $\abs{g}^2$ with the value in \eqref{eq:IRvalue} can only make it so.
Thus we conclude that the procedure we use here is at least valid at one-loop.

Most rigorously, we should be using the method introduced in \cite{Chester:2015wao} for the correct prescription.
This involves treating everything with explicit dependence on the RG scale, and tune the coupling so that the dependence on the arbitrary scale vanishes.
In our case, we would need to go beyond one-loop to see this, and because of the cumbersomeness we did not conduct this computation.

\subsection{Superfluid regime}

First we discuss the regime $Q\gg 16\pi^2/\abs{g\tau}^2$.
In the language of the last subsection, in this regime we have a large parameter $B=\mu\gg 1$.

\subsubsection{Contribution from $\phi_X$}

By expanding the summand $f_{\phi_X}(\ell)$ in the inverse $\ell$ expansion, we have
\begin{equation}
c_{5,0}=-\frac{5}{8}(\mu^2-1)^2,
\end{equation}
and also by inspection we can see that
\begin{equation}
K(\mu)\propto \mu^4+\text{(subleading)}.
\end{equation}
This can also be expected without calculation, from the EFT at large charge.
The form of $K(\mu)$ cannot the anything else -- otherwise it would not reproduce the semi-classical operator dimension at large charge, as expected from the EFT.
Also, we will not compute the coefficient in front of $\mu^4$, since the contribution is suppressed in comparison to $\mu^4\log \mu$.

By plugging these into the formula in the last subsection, we have
\begin{align}
\Delta_1^{\phi_X}&=
\frac{5}{8}\mu^4\log \mu+O(\mu^4)\\
&=\frac{5}{24}\log\left(\frac{\abs{g\tau}^2Q}{4m^3\alpha(D)}\right)\left(\frac{\abs{g\tau}^2Q}{4m^3\alpha(D)}\right)^{4/3}+O\left(\left(\frac{\abs{g\tau}^2Q}{4m^3\alpha(D)}\right)^{4/3}\right)
\end{align}

\subsubsection{Contribution from $\psi_X$}

As we have
\begin{equation}
c_{5,0}=\frac{1}{4}\mu^4+o(\mu^4),
\end{equation}
and
\begin{equation}
K(\mu)\propto \mu^4+\text{(subleading)},
\end{equation}
the sum can be computed to be
\begin{align}
\Delta_1^{\psi_X}&=
-\frac{1}{4}\mu^4\log \mu+O(\mu^4)\\
&=-\frac{1}{12}\log\left(\frac{\abs{g\tau}^2Q}{4m^3\alpha(D)}\right)\left(\frac{\abs{g\tau}^2Q}{4m^3\alpha(D)}\right)^{4/3}+O\left(\left(\frac{\abs{g\tau}^2Q}{4m^3\alpha(D)}\right)^{4/3}\right)
\end{align}

\subsubsection{Contribution from $\phi_{Y,Z}$}

As we have
\begin{equation}
c_{5,0}=-\frac{4+\abs{\tau}^2}{4\abs{\tau}^4}\mu^4+o(\mu^4),
\end{equation}
and
\begin{equation}
K(\mu)\propto \mu^4+\text{(subleading)},
\end{equation}
the sum can be computed to be
\begin{align}
\Delta_1^{\phi_{Y,Z}}&=
\frac{4+\abs{\tau}^2}{4\abs{\tau}^4}\mu^4\log \mu+O(\mu^4)\\
&=\frac{4+\abs{\tau}^2}{12\abs{\tau}^4}\log\left(\frac{\abs{g\tau}^2Q}{4m^3\alpha(D)}\right)\left(\frac{\abs{g\tau}^2Q}{4m^3\alpha(D)}\right)^{4/3}+O\left(\left(\frac{\abs{g\tau}^2Q}{4m^3\alpha(D)}\right)^{4/3}\right)
\end{align}

\subsubsection{Contribution from $\psi_{Y,Z}$}

As we have
\begin{equation}
c_{5,0}=-\frac{-2+\abs{\tau}^2}{2\abs{\tau}^4}\mu^4+o(\mu^4),
\end{equation}
and
\begin{equation}
K(\mu)\propto \mu^4+\text{(subleading)},
\end{equation}
the sum can be computed to be
\begin{align}
\Delta_1^{\phi_{Y,Z}}&=
\frac{-2+\abs{\tau}^2}{2\abs{\tau}^4}\mu^4\log \mu+O(\mu^4)\\
&=\frac{-2+\abs{\tau}^2}{6\abs{\tau}^4}\log\left(\frac{\abs{g\tau}^2Q}{4m^3\alpha(D)}\right)\left(\frac{\abs{g\tau}^2Q}{4m^3\alpha(D)}\right)^{4/3}+O\left(\left(\frac{\abs{g\tau}^2Q}{4m^3\alpha(D)}\right)^{4/3}\right)
\end{align}

\subsection{Weak-coupling Regime}

Next we study the regime $Q\ll 16\pi^2/\abs{g\tau}^2,\, 16\pi^2/\abs{g}^2$.
In this regime, we have no large parameter. We will calculate the sum in perturbation theory in $y=\mu-m\ll 1$.

\subsubsection{Contribution from $\phi_X$}

At order $y$ we have
\begin{align}
c_{1,0}&=1\\ c_{2,0}&=3\\ c_{3,0}&=3+2y+O(y^2)\\ c_{4,0}&=1+2y+O(y^2)\\ c_{5,0}&=0+O(y^2) \\
c_{5,1}&=-\frac{31}{120}-\frac{5}{6}y+O(y^2),
\end{align}
while
\begin{equation}
F(0)=f_{\phi_X}(0)=1+\frac{3}{2}y+O(y^2).
\end{equation}
Using the expansion \eqref{eq:small_Q_dimension} and since $\tilde{F}(\ell)=O(y^2)$, the sum becomes
\begin{align}
\Delta_1^{\phi_X}
=-\frac{1}{2}y=\textcolor{black}{m\left[-\frac{1}{2}\frac{\abs{g\tau}^2Q}{8m^3\alpha(D)}+O\left(\left(\frac{\abs{g\tau}^2Q}{8m^3\alpha(D)}\right)^2\right)\right]\;.}
\end{align}
where we plugged in the result \eqref{eq:A} for $\mu$.

\subsubsection{Contribution from $\psi_X$}

We have
\begin{align}
c_{1,0}&=-1\\ c_{2,0}&=-\frac{9}{2}\\ c_{3,0}&=-\frac{13}{2}-2y+O(y^2)\\ c_{4,0}&=-3-3y+O(y^2)\\ c_{5,0}&=0+O(y^2) \\
c_{5,1}&=\frac{29}{30}+\frac{11}{6}y+O(y^2),
\end{align}
while
\begin{equation}
F(0)=f_{\psi_X}(0)=-3-3y+O(y^2).
\end{equation}
Again using the expansion \eqref{eq:small_Q_dimension} and the fact that $\tilde{F}(\ell)=O(y^2)$, the sum becomes
\begin{align}
\Delta_1^{\psi_X}
=\frac{1}{2}y=\textcolor{black}{m\left[\frac{1}{2}\frac{\abs{g\tau}^2Q}{8m^3\alpha(D)}+O\left(\left(\frac{\abs{g\tau}^2Q}{8m^3\alpha(D)}\right)^2\right)\right]}
\end{align}

\subsubsection{Contribution from $\phi_{Y,Z}$}

We have
\begin{align}
c_{1,0}&=2\\ c_{2,0}&=6\\ c_{3,0}&=6+\frac{4}{\abs{\tau}^2}y+O(y^2)\\ c_{4,0}&=2+\frac{4}{\abs{\tau}^2}y+O(y^2)\\ c_{5,0}&=0+O(y^2) \\
c_{5,1}&=-\frac{31}{60}-\frac{5}{3\abs{\tau}^2}y+O(y^2),
\end{align}
while
\begin{equation}
F(0)=f_{\phi_{Y,Z}}(0)=2+\left(1+\frac{2}{\abs{\tau}^2}+\sqrt{1-\frac{5}{\abs{\tau}^2}+\frac{4}{\abs{\tau}^4}}\right)y+O(y^2).\end{equation}
Again we have $\tilde{F}(\ell)=O(y^2)$, and so the sum is
\begin{align}
\Delta_1^{\phi_{Y}}+\Delta_1^{\phi_{Z}}
&=\left(1-\frac{2}{\abs{\tau}^2}+\sqrt{1-\frac{5}{\abs{\tau}^2}+\frac{4}{\abs{\tau}^4}}\right)y
\\
&=\textcolor{black}{m\left[\left(1-\frac{2}{\abs{\tau}^2}+\sqrt{1-\frac{5}{\abs{\tau}^2}+\frac{4}{\abs{\tau}^4}}\right)\frac{\abs{g\tau}^2Q}{8m^3\alpha(D)}+O\left(\left(\frac{\abs{g\tau}^2Q}{8m^3\alpha(D)}\right)^2\right)\right]}\;.
\end{align}

\subsubsection{Contribution from $\psi_{Y,Z}$}

We have
\begin{align}
c_{1,0}&=-2\\ c_{2,0}&=-9\\ c_{3,0}&=-13-\frac{4}{\abs{\tau}^2}y+O(y^2)\\ c_{4,0}&=-6-\frac{6}{\abs{\tau}^2}y+O(y^2)\\ c_{5,0}&=0+O(y^2) \\
c_{5,1}&=\frac{29}{15}+\frac{11}{3\abs{\tau}^2}y+O(y^2),
\end{align}
while
\begin{equation}
F(0)=f_{\psi_{Y,Z}}(0)=-6-\frac{6}{\abs{\tau}^2}y+O(y^2).
\end{equation}
Again we have $\tilde{F}(\ell)=O(y^2)$, and the sum becomes
\begin{align}
\Delta_1^{\psi_{Y}}+\Delta_1^{\psi_{Z}}
=\frac{1}{\abs{\tau}^2}y=\textcolor{black}{m\left[\frac{1}{\abs{\tau}^2}\frac{\abs{g\tau}^2Q}{8m^3\alpha(D)}+O\left(\left(\frac{\abs{g\tau}^2Q}{8m^3\alpha(D)}\right)^2\right)\right]}
\end{align}

\subsection{Near-BPS regime}

Finally we study the regime $ 16\pi^2/\abs{g}^2 \ll Q \ll 16\pi^2/\abs{g\tau}^2$.
In this regime there is once again a large parameter $\abs{g}^2Q\gg 1$, but there is also a small parameter $\abs{g\tau}^2Q\ll 1$.
We will set 
\begin{equation}
y\equiv \mu-m\ll 1,\quad B=\left(\frac{y}{\abs{\tau}^2}\right)^{1/2}\gg 1,
\end{equation}
and apply the method explained in the last subsection.
We also explore the regime where $\abs{\tau}^2$ is the smallest parameter, so that $B\abs{\tau}^{p>0}\ll 1$.
In terms of $Q$, we have
\begin{equation}
y=\frac{\abs{g\tau}^2Q}{8m^3\alpha(D)},\quad B=\left(\frac{\abs{g}^2Q}{8m^3\alpha(D)}\right)^{1/2}
\end{equation}

\subsubsection{Contribution from $\phi_X$ and $\psi_X$}

As the contributions from $\phi_X$ and $\psi_X$ are the same as in Regime 2,
in total their contributions cancel with each other and gives
\begin{equation}
\Delta_1^{\phi_X}+\Delta_1^{\psi_X}=0
\end{equation}

\subsubsection{Contribution from $\phi_{Y,Z}$ and $\psi_{Y,Z}$}

We would like to evaluate the sum
\begin{equation}
\Delta_1^{\phi_{Y,Z}}+\Delta_1^{\psi_{Y,Z}}=\sum_{\ell}f(\ell)\equiv \sum_{\ell} \left(f_{\phi_{Y,Z}}(\ell)+f_{\psi_{Y,Z}}(\ell)\right).
\end{equation}
First let us expand the summand in the inverse $\ell$ expansion, to get the coefficient
\begin{equation}
c_{5,0}=-3B^4\abs{\tau}^2+O(B^6\abs{\tau}^4)
\end{equation}
Also as in region 1, we expect
\begin{equation}
K(B)=\sum_{j=0}^{\infty}\abs{\tau}^{2j}k_{j}(B^2),
\end{equation}
where $k_j(B)$ has a asymptotic expansion at large $B$ like
\begin{equation}
k_j(B^2)=\alpha_{j,4} B^4+ \alpha_{j,2} B^2+O(1).
\end{equation}

The final result becomes, therefore,
\begin{equation}
\Delta_1^{\phi_{Y,Z}}+\Delta_1^{\psi_{Y,Z}}=k_0(B^2)+\abs{\tau}^2\left(3B^4\log B+k_1(B^2)\right)+O\left(\abs{\tau}^4\right)
\end{equation}
Incidentally, as discussed in the main body of the text, $k_0(B)$ can be determined uniquely so that at $\tau=0$, the operator dimension reproduces the BPS bound.

\bibliographystyle{JHEP} 
\bibliography{references,cond-mat}

\end{document}